*Word Count: 12684*

Corresponding Author: Sune Nørhøj Jespersen.

Mail: sune@cfin.au.dk

# Larmor frequency shift from magnetized cylinders with arbitrary orientation distribution

Anders Dyhr Sandgaard[1], Noam Shemesh[2], Valerij G. Kiselev[3], Sune Nørhøj Jespersen[1,4]

[1]Center for Functionally Integrative Neuroscience, Department of Clinical Medicine, Aarhus University, Denmark

[2]Champalimaud Research, Champalimaud Centre for the Unknown, Lisbon, Portugal,

[3]Medical Physics, Department of Radiology, Faculty of Medicine, University of Freiburg, Freiburg, Germany

[4]Department of Physics and Astronomy, Aarhus University, Denmark

# Abstract

Magnetic susceptibility of tissue can provide valuable information about its chemical composition and microstructural organization. However, the relation between the magnetic microstructure and the measurable Larmor frequency shift is only understood for a few idealized cases. Here we analyze the microstructure formed by magnetized, NMR-invisible infinite cylinders suspended in an NMR-reporting fluid. Through simulations, we scrutinize various geometries of mesoscopic Lorentz cavities, inclusions, and show that the cavity size should be approximately one order of magnitude larger than the width of the inclusions. We also analytically derive the Larmor frequency shift for a population of cylinders with arbitrary orientation dispersion and show that it is determined by the $l = 2$ Laplace expansion coefficients $p_{2m}$ of the cylinders' orientation distribution function. Our work underscores the need to account for microstructural organization when estimating magnetic tissue properties.

## Introduction

The effect of tissue magnetic susceptibility on the MRI signal was initially considered an imaging artifact, but during the 1990s it was recognized as more than just a nuisance. One of its first uses, dubbed BOLD imaging, measured changes in oxygenation of red blood cells as a consequence of cerebral metabolic process[1–3]. The amount of deoxygenated blood is measurable as a signal phase for the reporting spins and as an increased transverse relaxation rate. Since then, many novel methods utilizing tissue-specific susceptibility have been developed[4–12], and have found use in neurological research.

With the increase of magnetic field strength in MRI scanners during the 2000s, a stronger contrast between gray (GM) and white matter (WM) tissue was clearly visible in both the phase and relaxation rate of a gradient echo[13–15]. The initial assumption was that differences in signal phase could be attributed to voxel-specific scalar magnetic susceptibility[16]. This ignited the hope of inverting such phase measurements to generate susceptibility maps of the brain, initiating the field of Quantitative Susceptibility Mapping (QSM)[17–26]. Experiments performed in ex-vivo tissue later demonstrated that the measured signal phase could not be fully explained by isotropic models[27]. Instead, experiments found evidence for magnetic anisotropy in WM[28–30]. As a consequence, QSM was extended to susceptibility tensors[28] (e.g., Susceptibility Tensor Imaging (STI)), whose determination typically requires rotations of the sample with respect to the main magnetic field.

However, the fundamental convolutional relationship between field offset and susceptibility exploited in conventional QSM and STI is only valid when the tissue behaves as a homogenously magnetized fluid. In this context, an important observation was the loss of phase contrast between WM axons parallel to the scanner field and cerebrospinal fluid (CSF). This effect was dubbed the WM-darkness-effect[33] and could not be explained directly by the observed WM susceptibility anisotropy. To explain this phenomena, He and Yablonskiy introduced a theoretical framework called the Generalized Lorentzian approach[31] (GLA), which included an explicit contribution to anisotropy from local magnetic microstructure. This was achieved with the use of a mesoscopic Lorentz cavity[31,34–37] emerging from subdividing the sample into a near and far region to compute a mesoscopically averaged Larmor frequency shift (see Figure 1 for overview). The size of the near region must be large enough to allow the far region to be described by averaged medium parameters, whereas the field in the near region is computed with account of explicit microstructure. This in turn defined a mesoscopic and macroscopic frequency contribution respectively, and could provide an explanation to the low phase contrast between WM and CSF. The GLA model was later extended to include an axially symmetric and uniform susceptibility tensor for WM, called the Generalized Lorentzian Tensor Approach[38] (GLTA), which provided a more complete description for QSM/STI, but is seldomly used in experiments. GLTA also considered frequency shifts from spherical sources with scalar susceptibility and exchange effects near the water-myelin interface to explain a discrepancy between frequency measurements inside and outside fixed rat optic nerve[39]. This experiment was later redone[40] in fresh porcine optic nerve to determine the frequency contribution from both magnetic and structural anisotropy of WM microstructure, and its impact on QSM and STI. While numerous phenomenological and modelling approaches[31,33,38,40–44] have been developed throughout the years to explain e.g. how the Larmor frequency depends on both susceptibility and microstructure, it has been limited to parallel or isotropically dispersed cylinders and/or low volume fraction. In recent years, Ruh, Scherer and Kiselev have shown that the frequency shift from a uniformly magnetized microstructure can be described for arbitrary geometries in terms of structural autocorrelations[36,45]. Their framework provides a general analytical background for the Lorentzian correction in GLTA, and a recipe for future model development that goes beyond common limitations mentioned earlier. This was demonstrated in their papers by providing a theoretical result for parallel solid cylinders with arbitrary volume fraction. However, to this day, there is no quantitative result for cylinders with arbitrary orientation dispersion and volume fraction.

In this work, we revisit how the Larmor frequency, measured by NMR and MRI, depends on this magnetic microstructure. While such analyses have been considered previously[38,43,45], we start at the microscopic level and use the principle of coarse-graining to show how the measured frequency relates to a mesoscopic average of microscopic Larmor frequencies and depends on explicit magnetic microstructure within a mesoscopic Lorentz cavity. Furthermore, we show with simulations that sensitivity to local magnetic microstructure can be captured with high accuracy by a mesoscopic Lorentz cavity an order of magnitude greater than the microstructural correlation length. The error incurred with this construction decreases as a power law of the size of the mesoscopic cavity with an exponent depending on the microstructural universality class[46]. Using our framework, we consider the Larmor frequency from a population of solid cylinders, whose length is much greater than their diameter, with a scalar susceptibility different from the surrounding media. We derive how it depends on the $l = 2$ Laplace expansion coefficients $p_{2m}$ of the cylinder (fiber) orientation distribution (fODF) and validate it with simulations.

While our model is simple compared to actual tissue microstructure, we believe it represents an important step for understanding the effects of orientation dispersion for e.g., blood vessels or for axons in WM. Thus, if orientation dispersion can be incorporated into a biophysical model of magnetic field effects, dMRI could be used to quantify the level of dispersion, and in the end, provide better susceptibility estimations. Vice versa, susceptibility effects could also be a potential candidate to constrain the estimation of diffusion related parameters[47–49].

# Theory

We start by outlining the system of consideration along with the central theoretical aspects of the local microscopic Larmor frequency shift[38,43,45]. Second, we consider the coarse-grained Larmor frequency and split it into contributions from nearby and distant sources, by employing the mesoscopic Lorentz cavity to quantify the contribution from microscopic and macroscopic length scales individually. Third, using the coarse-grained Larmor frequency we describe the MRI measured Larmor frequency. Last, we apply the framework to a specific system consisting of infinitely long solid cylinders with an arbitrary orientation distribution to bridge the theoretical gap between ideally parallel or maximally dispersed cylinders and understand its impact on susceptibility estimation.

*System of consideration*

We describe the macroscopic sample of volume $\mathrm{V}$ as a porous medium of impermeable microscopic magnetic inclusions immersed in an NMR-visible fluid with magnetic susceptibility $\chi^W$. The spatial organization of the microscopic inclusions is encapsulated in the microscopic indicator function $v(\boldsymbol{r})$, which is 1 inside inclusions and 0 otherwise (as depicted in Figure 1). We assume inclusions are weakly dia- or paramagnetic, and uniformly magnetized along the external field $\mathbf{B}_0 = \mathrm{B}_0\hat{\mathbf{B}}$, where $\hat{\mathbf{B}}$ is a unit vector (as all hatted vectors in what follows). This magnetization response is characterized by a microscopic (cellular or molecular level) magnetic susceptibility $\chi(\boldsymbol{r})$ ($|\chi| << 1$, and is given relative to the susceptibility $\chi^W$ of the NMR fluid), which contains all the details of the considered medium. For porous media embedded in a single liquid, it is convenient to include $v(\boldsymbol{r})$ in $\chi(\boldsymbol{r})$, as susceptibility is zero outside inclusions. In general, $\chi(\boldsymbol{r})$ is a tensor, but we focus this study on scalar $\chi(\boldsymbol{r})$.

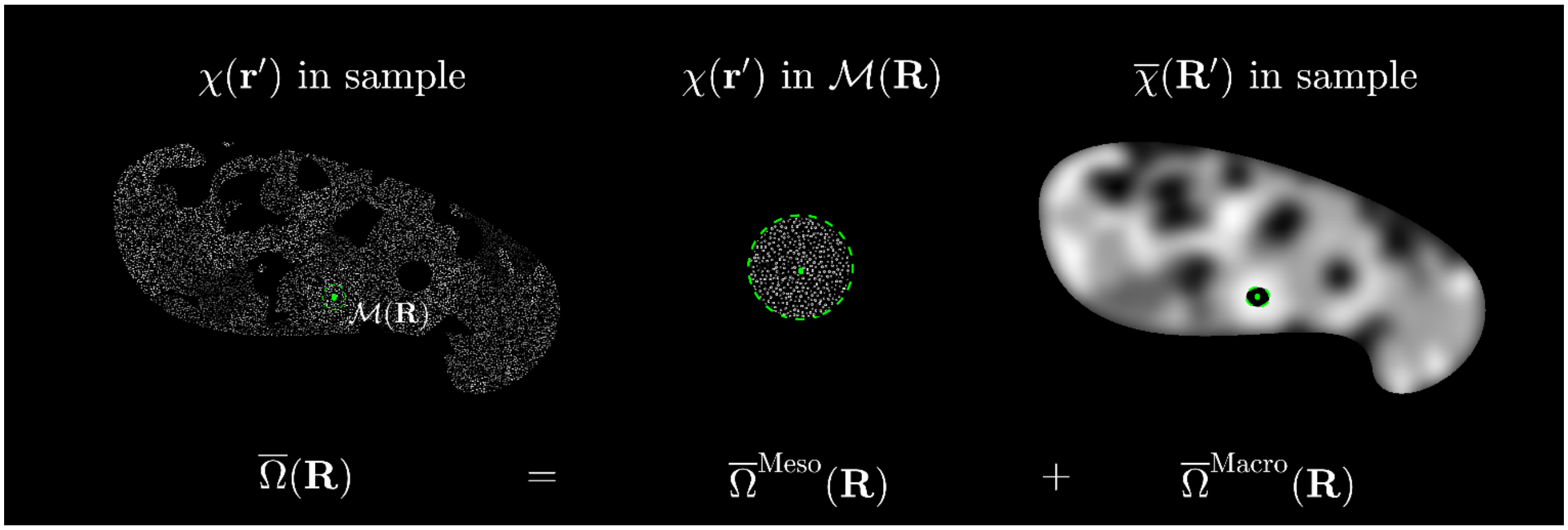

*Figure 1 - Magnetic microstructure model. Local magnetic susceptibility $\chi(\boldsymbol{r}')$ from multiple regions with an associated microstructure. The coarse-grained Larmor frequency $\overline{\Omega}(\mathbf{R})$ is given by the microscopic Larmor frequency $\Omega(\boldsymbol{r})$ felt by a spin at position $\boldsymbol{r}$ averaged within a mesoscopic sphere $\mathrm{M}(\mathbf{R})$. $\overline{\Omega}(\mathbf{R})$ can be described by two contributions:*

*First contribution is from explicit microstructure within $\mathrm{M}(\mathbf{R})$, which defines the mesoscopic Larmor frequency $\bar{\Omega}^{\mathrm{Meso}}(\mathbf{R})$. The second depends on the Larmor frequency at $\mathbf{R}$ induced by the mesoscopic averaged magnetic microstructure $\bar{\chi}(\mathbf{R}')$. This defines the local macroscopic Larmor frequency $\bar{\Omega}^{\mathrm{Macro}}(\mathbf{R})$. The image depicts a mesoscopic spherical cavity in $\bar{\Omega}^{\mathrm{Macro}}(\mathbf{R})$ (size may be exagerated). Due to the functional form of the dipole field, Eq. (2) , we can include the mesoscopic sphere in the integration when computing $\bar{\Omega}^{\mathrm{Macro}}(\mathbf{R})$.*

*Local Larmor frequency $\Omega(\boldsymbol{r})$*

When exposed to a magnetic field $\mathbf{B}_0 = \mathrm{B}_0\hat{\mathbf{B}}$, the inclusions become magnetized and produce a secondary magnetic field $\Delta\mathbf{B}(\boldsymbol{r})$, which perturbs the local Larmor frequency by $\Omega(\boldsymbol{r})$. To the first order in $\chi$, $\Omega(\boldsymbol{r}) = \gamma\hat{\mathbf{B}}^{\mathrm{T}}\Delta\mathbf{B}(\boldsymbol{r})$, which is a reasonable approximation when $|\chi| \sim \mathrm{ppm}$ [23]. Here $\gamma$ defines the fluid-specific gyromagnetic ratio. The frequency shift $\Omega(\boldsymbol{r})$ can be written as a convolution over the sample:

$$\Omega(\boldsymbol{r}) = \gamma \mathrm{B}_0 \hat{\mathbf{B}}^{\mathrm{T}} \int d\boldsymbol{r}'\, \boldsymbol{\Upsilon}(\boldsymbol{r}-\boldsymbol{r}')\,\chi(\boldsymbol{r}')\,\hat{\mathbf{B}}. \quad (1)$$

$\boldsymbol{\Upsilon}(\boldsymbol{r})$ defines the dipole field outside an infinitesimal sphere of radius $\varepsilon$ (in SI units)

$$\boldsymbol{\Upsilon}(\boldsymbol{r}) = \frac{1}{4\pi}\left(\frac{3\hat{\mathbf{r}}\hat{\mathbf{r}}^{\mathrm{T}} - \mathbf{I}}{r^3}\right),\ r > \varepsilon. \quad (2)$$

Since there are no field-inducing sources at the measurement point $\boldsymbol{r}$, the value at $r < \varepsilon$ can so far be chosen at our convenience. For reasons which become apparent in the next sections, we thus set $\boldsymbol{\Upsilon}(\boldsymbol{r}) = 0$, for $r < \varepsilon$. The functional form of $\boldsymbol{\Upsilon}(\boldsymbol{r})$ now corresponds to the sphere of Lorentz-corrected elementary dipole-field[50]. In Fourier space, $\boldsymbol{\Upsilon}(\boldsymbol{k})$ is

$$\boldsymbol{\Upsilon}(\boldsymbol{k}) = \frac{1}{3}\mathbf{I} - \hat{\boldsymbol{k}}\hat{\boldsymbol{k}}^{\mathrm{T}},\ k > 0, \quad (3)$$

and zero for $k = 0$, where $\boldsymbol{k}$ is the wavevector.

Since Eq. (1) describes the frequency shift from the susceptibility difference $\chi(\boldsymbol{r})$ with respect to the MR fluid, we get an additional frequency shift that depends only on the fluid susceptibility $\chi^W$ and sample shape

$$\Omega^{W}(\boldsymbol{r}) = \chi^{W}\gamma B_0 \hat{\mathbf{B}}^{\mathrm{T}} \mathbf{N}^{W}(\boldsymbol{r}) \hat{\mathbf{B}}. \tag{4}$$

Here $\mathbf{N}^{W}(\boldsymbol{r})$ defines the sample-specific Lorentz-corrected point-demagnetization tensor[51,52]

$$\mathbf{N}^{W}(\boldsymbol{r}) = \int d\boldsymbol{r}' \, \boldsymbol{\Upsilon}(\boldsymbol{r}-\boldsymbol{r}'), \tag{5}$$

where the integration is understood to be over the sample. Here, $\boldsymbol{\Upsilon}(\boldsymbol{r})$ must be the Lorentz-corrected dipole-field (cf. Eq. (3)) due to averaging across the many molecular degrees of freedom associated with molecular dynamics of the NMR fluid[50].

In the next sections we focus on the microscopic relation $\Omega(\boldsymbol{r})$, Eq. (1), and outline how a similar "coarse-graining" must also be carried out across its microscopic degrees of freedom to reach a description of the MR measured Larmor frequency.

*Sample microstructure and coarse-graining*

In MRI, the nominal imaging resolution is measured in millimeters, while the size of inclusions (biological cells) is typically micrometers. This separation of scales introduces a substantial loss of information, as we do not observe single microscopic inclusions, but rather the effects of a myriad of them. This enormous loss of sensitivity to the microstructural details can be described by *coarse-graining*, which is the massive averaging inherent to the macroscopic measurement. While this phenomenon is general, we illustrate it in the present context of phase MRI. Consider the genuine microscopic Larmor frequency shift, Eq. (1). Since it is microscopic, only its coarse-grained version, the locally (voxel) averaged frequency $\bar{\Omega}(\mathbf{R})$, is practically measurable. In this case, the coarse-graining is nothing but filtering $\Omega(\boldsymbol{r})$ with a kernel, $f(\boldsymbol{r})$, that describes the averaging effect introduced by a natural process such as diffusion[53] or the measurement itself such as MR imaging,

$$\bar{\Omega}(\mathbf{R}) = \int d\boldsymbol{r} \, f(\boldsymbol{r}) \Omega(\mathbf{R}-\mathbf{r}), \quad \int d\boldsymbol{r} f(\boldsymbol{r}) = 1. \tag{6}$$

To look at it in more detail, we analyze it in Fourier space, where $\bar{\Omega}(\mathbf{R})$ becomes

$$\bar{\Omega}(\boldsymbol{k}) = f(\boldsymbol{k}) \Omega(\boldsymbol{k}). \tag{7}$$

The spectral power of $\Omega(\boldsymbol{k})$ is distributed from very large $\boldsymbol{k}$, corresponding to the microstructure, to small $\boldsymbol{k}$ describing variations over the macroscopic scale such as the brain. In contrast, the spectral power of $f(\boldsymbol{k})$ is present for small $\boldsymbol{k}$ only. For instance, diffusion in the long-time regime results in a nearly Gaussian shape of $f(\boldsymbol{k})$ with the width equal the inverse diffusion length. For imaging, $f(\boldsymbol{k})$ is the Fourier transformed point-spread-function (PSF) with an overall width given by the inverse spatial resolution, which is much smaller than the inverse cell size. Therefore, $\bar{\Omega}(\boldsymbol{k})$ is largely insensitive to the form of $\Omega(\boldsymbol{k})$ at high spatial frequencies leaving only the sensitivity to the low wavenumbers. Note that coarse-graining is a well-known physical concept used also in NMR and MRI where the diffusion propagator acts as a time-dependent Gaussian filter, increasingly homogenizing the microstructure with an imprint on diffusion[46,54] and transverse relaxation[55].

We refer to the low $\boldsymbol{k}$ domain and the corresponding large distances as the *macroscopic scale*, which is typically set by the voxel or the sample size if measured as a whole in a spectrometer. The other limit is the *microscopic scale* associated with the cell size or below. In view of the large separation between these scales, one can define an intermediate *mesoscopic scale* at which the individual cellular features are already well averaged giving rise to local statistical characteristics. On the other hand, this spatial scale is finer than the macroscopic one. A more precise understanding of the magnitude of this scale is one of the aims of this study.

The relationship between the microscopic Larmor frequency and the magnetic susceptibility in Eq. (1) does not amount to coarse-graining. While it looks similar to Eq. (4), the difference is the form of the "filter", which in this case is the dipole kernel, Eqs. (2) and (3). The dipole kernel is at its power-law form and therefore scale invariant, which means that it does not introduce any new scale. As it scales as $\boldsymbol{r}^{-3}$, the Larmor frequency depends on both microscopic and macroscopic scales on an equal footing. Had it been slower, the macroscopic scale would dominate, while a more rapid decay would favor the microscopic. The above-mentioned mesoscopic Lorentz cavity[38,43,45] is a way to separate these scales for identifying the relevant, measurable medium parameters as discussed below.

While the term *microstructure* is used to describe the complex tissue architecture at the microscopic scale (micrometer, the cell size in the biological context), we can similarly define the *magnetic microstructure* as a detailed description of magnetic properties on this scale, characterized by the magnetic susceptibility and positions of the inclusions. Such a detailed explicit description is unsurmountable for any practical purpose. A tractable description operates instead with statistical properties of the medium that crystallize at the mesoscopic level beyond which the coarse-graining becomes effective. As the microscopic $\chi(\boldsymbol{r})$ contains $v(\boldsymbol{r})$, magnetic microstructure depends on the structural architecture. In the simplest case as studied here, they differ by a coefficient, $\chi(\boldsymbol{r})=\chi v(\boldsymbol{r})$ [33,45], but one can include finer structural details as the radial arrangement of the susceptibility tensor in the myelin layers[56], resulting in nontrivial macroscopic (coarse-grained) tissue properties[30,39,57].

*Coarse-grained Larmor frequency* $\bar{\Omega}(\mathbf{R})$

To connect to macroscopic quantities measurable by the NMR and MRI Larmor frequency we must consider the coarse-grained Larmor frequency $\bar{\Omega}(\mathbf{R})$ defined as the mesoscopic average of microscopic Larmor frequencies $\Omega(\boldsymbol{r})$

$$\bar{\Omega}(\mathbf{R})=\frac{1}{\left(1-\bar{\zeta}(\mathbf{R})\right)|\mathrm{M}|}\int_{\mathrm{M}(\mathbf{R})} d\boldsymbol{r}\left(1-v(\boldsymbol{r})\right)\Omega(\boldsymbol{r}). \tag{8}$$

The factor $\bar{\zeta}(\mathbf{R})$ is the volume fraction of the NMR invisible inclusions

$$\bar{\zeta}(\mathbf{R})=\frac{1}{|\mathrm{M}|}\int_{\mathrm{M}(\mathbf{R})} d\boldsymbol{r}\, v(\boldsymbol{r}), \tag{9}$$

while $1-\bar{\zeta}(\mathbf{R})$ is the volume fraction of the NMR visible fluid. Comparing to Eq. (4), $\bar{\Omega}(\mathbf{R})$ is given by a convolution between $\Omega(\boldsymbol{r})$ and the indicator function of a mesoscopic sphere $\mathrm{M}(\mathbf{R})$ centered at $\mathbf{R}$ of volume $|\mathrm{M}|$. In principle, $\mathrm{M}(\mathbf{R})$ can be chosen as having any shape (validated in Simulation (a)), but we choose a sphere for later convenience (illustrated in Figure 1). Due to the $\boldsymbol{r}^{-3}$ scaling of the dipole field[50], details of magnetic microstructure become less relevant as $|\boldsymbol{r}-\boldsymbol{r}'|$ in Eq. (1) increases. As outlined in previous work[31,34–37]: when the magnetic microstructure is statistically homogenous near $\mathbf{R}$, it is convenient to decompose the sample into a near and far region. For this we use the mesoscopic sphere $\mathrm{M}(\mathbf{R})$ so the magnetic microstructure outside $\mathrm{M}(\mathbf{R})$ can be described by coarse-grained medium parameters (we investigate the effect of its size in simulation (a)). Hence, we keep explicit magnetic microstructure $\chi(\boldsymbol{r})$ within $\mathrm{M}(\mathbf{R})$, while outside $\mathrm{M}(\mathbf{R})$, magnetic microstructure is characterized by the locally averaged magnetic microstructure $\bar{\chi}(\mathbf{R}')$

$$\bar{\chi}(\mathbf{R}') = \frac{1}{|\mathrm{M}|} \int_{\mathrm{M}(\mathbf{R}')} d\boldsymbol{r}' \chi(\boldsymbol{r}'), \tag{10}$$

where we used $\mathrm{M}(\mathbf{R})$ due to the freedom in selecting the coarse-graining filter (validated in simulation (a)). Figure 1 illustrates this decomposition. Then, $\bar{\Omega}(\mathbf{R})$ can be decomposed into a mesoscopic contribution, $\bar{\Omega}^{\mathrm{Meso}}(\mathbf{R})$, depending on explicit magnetic microstructure inside $\mathrm{M}(\mathbf{R})$, and a macroscopic contribution, $\bar{\Omega}^{\mathrm{Macro}}(\mathbf{R})$, depending on coarse-grained magnetic microstructure outside $\mathrm{M}(\mathbf{R})$

$$\bar{\Omega}(\mathbf{R}) = \bar{\Omega}^{\mathrm{Meso}}(\mathbf{R}) + \bar{\Omega}^{\mathrm{Macro}}(\mathbf{R}). \tag{11}$$

The macroscopic contribution $\bar{\Omega}^{\mathrm{Macro}}(\mathbf{R})$ to the coarse-grained Larmor frequency $\bar{\Omega}(\mathbf{R})$ is given by the Larmor frequency at $\mathbf{R}$ induced by the coarse-grained susceptibility $\bar{\chi}(\mathbf{R})$

$$\bar{\Omega}^{\mathrm{Macro}}(\mathbf{R}) \approx \gamma \mathrm{B}_0 \hat{\mathbf{B}}^{\mathrm{T}} \int_{\mathbf{R}' \notin \mathrm{M}(\mathbf{R})} d\mathbf{R}' \, \Upsilon(\mathbf{R} - \mathbf{R}') \bar{\chi}(\mathbf{R}') \hat{\mathbf{B}}. \tag{12}$$

The larger the scale separation micro-meso-macro is, the more exact Eq. (10) becomes. By choosing $\mathrm{M}(\mathbf{R})$ as a sphere and the functional form of $\Upsilon(\mathbf{R})$ (cf. Eq. (3)), we may extend the integration in Eq. (12) to the entire sample, since integrating over $\mathrm{M}(\mathbf{R})$ yields zero.

The mesoscopic contribution $\bar{\Omega}^{\mathrm{Meso}}(\mathbf{R})$ to the coarse-grained Larmor frequency, $\bar{\Omega}(\mathbf{R})$, captures the induced frequency shift from sources within $\mathrm{M}(\mathbf{R})$

$$\bar{\Omega}^{\mathrm{Meso}}(\mathbf{R}) = \gamma \mathrm{B}_0 \hat{\mathbf{B}}^{\mathrm{T}} \frac{1}{\left(1 - \bar{\zeta}(\mathbf{R})\right)|\mathrm{M}|} \int_{\mathrm{M}(\mathbf{R})} d\boldsymbol{r} \left(1 - v(\boldsymbol{r})\right) \int_{\mathrm{M}(\mathbf{R})} d\boldsymbol{r}' \, \Upsilon(\boldsymbol{r} - \boldsymbol{r}') \chi(\boldsymbol{r}') \hat{\mathbf{B}}. \tag{13}$$

As outlined in previous work[38,45], $\bar{\Omega}^{\text{Meso}}(\mathbf{R})$ can be written in terms of a mesoscopic Lorentz tensor $\mathbf{L}(\mathbf{R})$. For an arbitrary magnetic microstructure (while we limit this study to scalar susceptibility, it is also valid for tensor susceptibility)

$$\bar{\Omega}^{\text{Meso}}(\mathbf{R}) = \gamma \text{B}_0 \hat{\mathbf{B}}^{\text{T}} \mathbf{L}(\mathbf{R}) \hat{\mathbf{B}}, \tag{14}$$

where

$$\mathbf{L}(\mathbf{R}) = -\frac{1}{\left(1-\bar{\zeta}(\mathbf{R})\right)} \int \frac{d\boldsymbol{k}}{(2\pi)^3} \boldsymbol{\Upsilon}(\boldsymbol{k}) \Gamma^{v\chi}(\boldsymbol{k};\mathbf{R}). \tag{15}$$

$\mathbf{L}(\mathbf{R})$ depends[45] on the cross-correlation function $\Gamma^{v\chi}$, whose generic form in Fourier space is

$$\Gamma^{v\chi}(\boldsymbol{k}) = \frac{v(\boldsymbol{k})\chi(\text{-}\boldsymbol{k})}{|\text{M}|}, \; k > 0, \tag{16}$$

and zero at $k = 0$. In Eq. (13), we wrote $\Gamma^{v\chi}(\boldsymbol{k};\mathbf{R})$ to indicate correlations within $\text{M}(\mathbf{R})$, whose size should be much larger than the correlation length of the magnetic microstructure. In the case of a sample with scalar susceptibility $\chi$, $\chi(\boldsymbol{r}) = \chi v(\boldsymbol{r})$, and the mesoscopic Lorentz tensor in this statistically homogenous case, simplifies to

$$\mathbf{L}(\mathbf{R}) = -\chi \mathbf{N}(\mathbf{R}), \quad \left(\chi(\boldsymbol{r}) = \chi v(\boldsymbol{r})\right), \tag{17}$$

where $\mathbf{N}(\mathbf{R})$ is a mesoscopic Lorentz-corrected magnetometric demagnetization tensor[51,52,58,59], depending only on microstructural correlations $\Gamma^{vv}$:

$$\mathbf{N}(\mathbf{R}) = \frac{1}{\left(1-\bar{\zeta}(\mathbf{R})\right)} \int \frac{d\boldsymbol{k}}{(2\pi)^3} \boldsymbol{\Upsilon}(\boldsymbol{k}) \Gamma^{vv}(\boldsymbol{k};\mathbf{R}), \tag{18}$$

$$\Gamma^{vv}(\boldsymbol{k}) = \frac{v(\boldsymbol{k})v(\text{-}\boldsymbol{k})}{|\text{M}|}, \, k > 0, \tag{19}$$

and zero for $k = 0$. Then, the total mesoscopic contribution to the coarse-grained Larmor frequency simplifies to

$$\bar{\Omega}^{\text{Meso}}(\mathbf{R}) = -\gamma\chi \text{B}_0 \hat{\mathbf{B}}^{\text{T}} \mathbf{N}(\mathbf{R}) \hat{\mathbf{B}}, \quad \left(\chi(\boldsymbol{r}) = \chi v(\boldsymbol{r})\right). \tag{20}$$

Here $\chi$ contributes one magnetic degree of freedom, while 5 additional structural (i.e., via $\Gamma^{\nu\nu}$ ) degrees of freedom characterize the symmetric and trace-free rank-2 tensor $\mathbf{N}(\mathbf{R})$ (as $\Upsilon(\boldsymbol{k})$ is traceless, c.f. Eq. (3)). Due to the $r^{-3}$ scaling of the dipole field, the averaged Larmor frequency $\bar{\Omega}(\mathbf{R})$ depends not only on coarse-grained magnetic susceptibility $\bar{\chi}(\mathbf{R})$ through the macroscopic contribution $\bar{\Omega}^{\text{Macro}}(\mathbf{R})$ , Eq. (10), but also explicit magnetic microstructure though $\bar{\Omega}^{\text{Meso}}(\mathbf{R})$, Eq. (12), which is not accounted for in QSM[17,18,23]. The demagnetization tensor $\mathbf{N}$, Eq. (16), has been considered in many previous studies[38,58–61], where solutions to simple structures such as a cylinder or sphere have been presented.

The coarse-grained frequency from pure water $\bar{\Omega}^{W}(\mathbf{R})$ can be written in similar form as $\bar{\Omega}^{\text{Macro}}(\mathbf{R})$ but with the susceptibility $\chi^{W}$ instead. Since the fluid-filled sample is uniform on the mesoscopic scale, it coincides with the microscopic frequency $\bar{\Omega}^{W}(\mathbf{R}) = \Omega^{W}(\mathbf{R})$ (cf. Eq. (4)).

*MRI measured Larmor frequency* $\bar{\Omega}_{\text{MRI}}(\mathbf{R})$

In appendix A, we derive how the NMR and MRI Larmor frequencies, measured in either the static or motional averaged regime, can be expressed through the coarse-grained Larmor frequency, Eq. (6), through Eqs. (10) and (12). Here we find for the MRI measured Larmor frequency, $\bar{\Omega}_{\text{MRI}}(\mathbf{R})$, at discrete sample positions $\mathbf{R}$ in image space:

$$\begin{aligned}\bar{\Omega}_{\text{MRI}}(\mathbf{R}) &\approx \bar{\Omega}^{\text{Meso}}(\mathbf{R}) + \bar{\Omega}^{\text{Macro}}(\mathbf{R}) + \bar{\Omega}^{W}(\mathbf{R}) \\ &\approx \gamma \mathrm{B}_0 \hat{\mathbf{B}}^{\mathrm{T}} \left( \mathbf{L}(\mathbf{R}) + \sum_{\mathbf{R}'} \bar{\Upsilon}(\mathbf{R}-\mathbf{R}')\bar{\chi}(\mathbf{R}') + \chi^{W}\mathbf{N}^{W}(\mathbf{R}) \right) \hat{\mathbf{B}}, \quad (\text{Slowly varying microstructure}).\end{aligned} \tag{21}$$

$\bar{\Upsilon}$ denotes the macrosopically averaged dipole field, Eq. (56) derived in appendix A2. As illustrated in our simulations (simulation (b)), the accuracy of Eq. (19) is high when the coarse-grained Larmor frequency $\bar{\Omega}(\mathbf{R})$ and susceptibility $\bar{\chi}(\mathbf{R}')$ vary slowly on the macroscopic scale. Keeping only the second term, $\bar{\Omega}^{\text{Macro}}(\mathbf{R})$, in Eq. (19) resembles the Larmor frequency considered in QSM[17,18] . However, Eq. (19) also differs from QSM, where the elementary dipole kernel $\Upsilon$, Eq. (3), is used instead of the averaged dipole kernel $\bar{\Upsilon}$, Eq. (56), hence neglecting the effects of imaging due to finite sampling. In simulation (c) we demonstrate the error from both neglecting the mesoscopic contribution, and using the elementary dipole field, Eq. (2) instead of the voxel averaged Eq. (56), when calculating $\bar{\Omega}^{\text{Macro}}(\mathbf{R})$.

To illustrate the theory developed above, we next consider $\mathbf{N}$, Eq. (16), for a specific microstructure.

*Larmor frequency of infinite solid cylinders with arbitrary orientation dispersion*

In this section we apply our framework (Eq. (16)) to a specific system consisting of a population of randomly placed infinitely long solid cylinders with scalar susceptibility, and with arbitrary orientation dispersion and radii, as illustrated in Figure 2. The derivation will mainly be qualitative, while a full analytical derivation of the simplified passages can be found in appendix B.

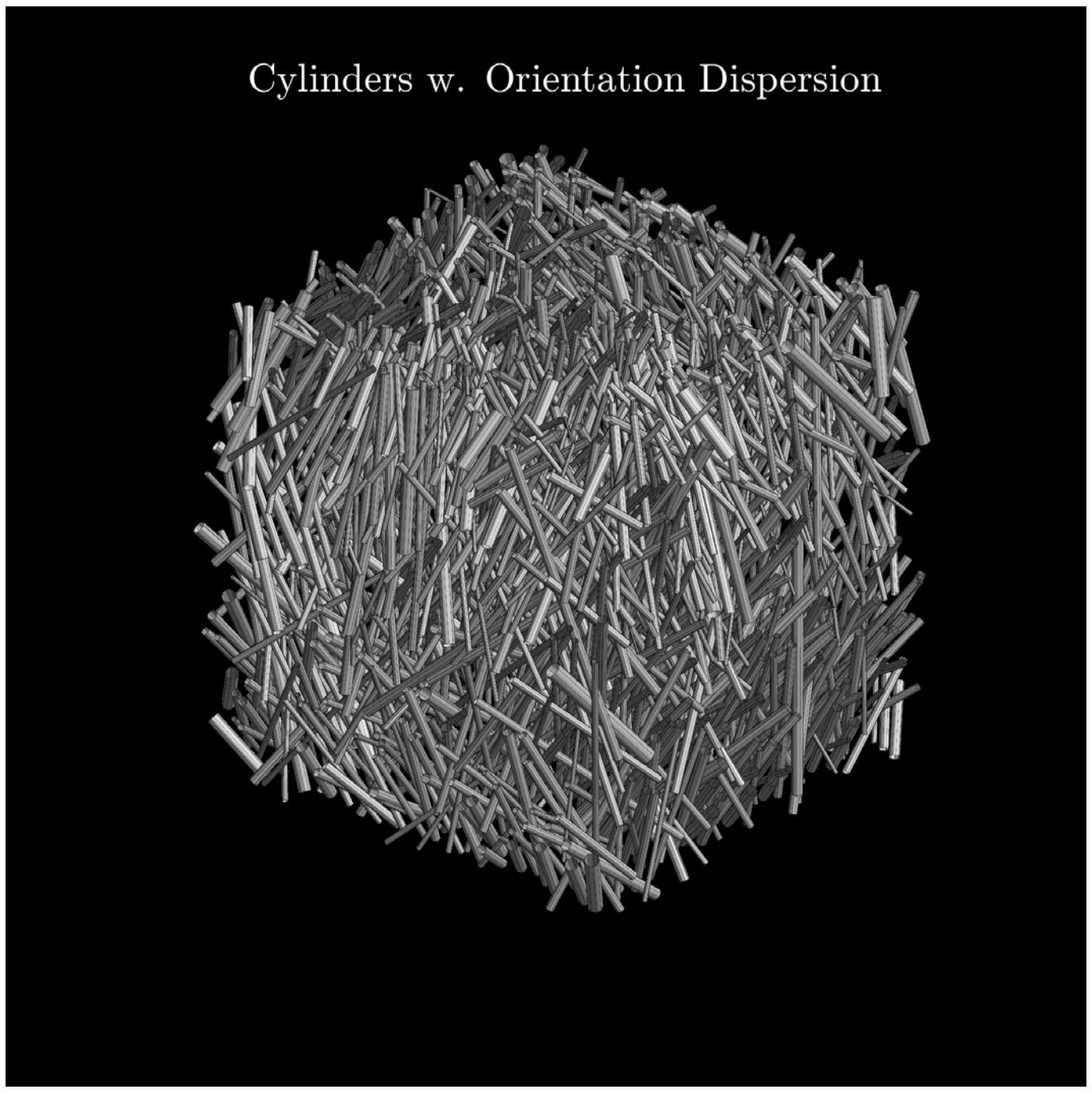

*Figure 2 - Microstructure Model. The system is modelled as a population of solid, infinite, and impermeable cylinders exhibiting orientation dispersion. Cylinders are randomly positioned, and their positions, orientations and radii are all independent.*

*Correlation function for infinitely long cylinders with orientation dispersion*

Consider now N infinitely long cylinders with volume fractions $\zeta_m$, where $m$ denotes the $m$'th cylinder. The total volume fraction becomes $\zeta = \sum_m \zeta_m$. As each cylinder has a unique orientation $\hat{\boldsymbol{n}}_m$ assumed to be statistically independent of its size, the total volume fraction is for later convenience also written as $\zeta = N\zeta_c$, where $\zeta_c = \langle \zeta_m \rangle$ is the average volume fraction. The structural correlation function $\Gamma^{vv}$ for such a system can be split into a sum over the autocorrelation $\Gamma^{\mathrm{Auto}}$ and cross-correlation $\Gamma^{\mathrm{Cross}}$

$$\Gamma^{vv}(\boldsymbol{k}) = \Gamma^{\mathrm{Auto}}(\boldsymbol{k}) + \Gamma^{\mathrm{Cross}}(\boldsymbol{k}), \tag{22}$$

where

$$\Gamma^{\mathrm{Auto}}(\boldsymbol{k}) = \sum_m \Gamma_m(\boldsymbol{k}), \tag{23}$$

and

$$\Gamma^{\mathrm{Cross}}(\boldsymbol{k}) = \sum_{m \neq w} \Gamma_{mw}(\boldsymbol{k}). \tag{24}$$

The mesoscopic demagnetization tensor, Eq. (16), is a sum of contributions associated with autocorrelation and cross-correlation, respectively

$$\mathbf{N} = \mathbf{N}^{\mathrm{Auto}} + \mathbf{N}^{\mathrm{Cross}}, \tag{25}$$

where each term in Eq. (23) can be written as a sum, like Eqs. (21)-(22), using

$$\mathbf{N}_m = \int \frac{d\boldsymbol{k}}{(2\pi)^3} \frac{\Gamma_m(\boldsymbol{k})}{(1-\zeta)} \Upsilon(\boldsymbol{k}), \tag{26}$$

$$\mathbf{N}_{mw} = \int \frac{d\boldsymbol{k}}{(2\pi)^3} \frac{\Gamma_{mw}(\boldsymbol{k})}{(1-\zeta)} \Upsilon(\boldsymbol{k}). \tag{27}$$

We now consider each contribution in turn.

*Mesoscopic contribution from autocorrelation*

The Fourier space indicator function for each cylinder, in the asymptotic limit where its length tends to infinity, positioned at $\hat{\boldsymbol{u}}$, and oriented along $\hat{\boldsymbol{n}}$, is found in appendix A1 Eq. (65)

$$v(\boldsymbol{k}) = e^{iku\hat{\boldsymbol{k}}\cdot\hat{\boldsymbol{u}}}\, v^{\mathrm{2D}}(k)\,\delta(\boldsymbol{k}\cdot\hat{\boldsymbol{n}}). \tag{28}$$

$v^{\mathrm{2D}}(k)$ is the indicator function of the cylinder perpendicular to its orientation $\hat{\boldsymbol{n}}$. The auto correlation function thus becomes (cf. Eq. (66))

$$\Gamma_m(\boldsymbol{k}) = 2\pi\,\Gamma_m^{\mathrm{2D}}(k)\,\delta(\boldsymbol{k}\cdot\hat{\boldsymbol{n}}_m). \tag{29}$$

The contribution to the mesoscopic dipole tensor $\mathbf{N}_m$ in Eq. (24) becomes

$$\mathbf{N}_m = \frac{1}{(1-\zeta)}\int\frac{d\boldsymbol{k}}{(2\pi)^2}\Gamma_m^{\mathrm{2D}}(k)\boldsymbol{\Upsilon}(\boldsymbol{k})\,\delta(\boldsymbol{k}\cdot\hat{\boldsymbol{n}}_m). \tag{30}$$

The delta function $\delta(\boldsymbol{k}\cdot\hat{\boldsymbol{n}}_m)$ defines a 2D polar integration perpendicular to the orientation $\hat{\boldsymbol{n}}(m)$ of the m'th cylinder. Using the axial symmetry of the correlation function and the fact that $\boldsymbol{\Upsilon}(\boldsymbol{k})$ does not depend on the radial distance $k$, the radial integral of Eq. (28) becomes

$$\frac{1}{1-\zeta}\int\frac{dkk}{(2\pi)^2}\Gamma_m^{\mathrm{2D}}(k) = \frac{1}{1-\zeta}\Gamma_m^{\mathrm{2D}}(\boldsymbol{r}=0) = \zeta_m, \tag{31}$$

while the angular integration yields

$$\int\frac{d\hat{\boldsymbol{k}}}{2\pi}\boldsymbol{\Upsilon}(\hat{\boldsymbol{k}})\,\delta(\hat{\boldsymbol{k}}\cdot\hat{\boldsymbol{n}}_m) = \frac{1}{3}\mathbf{I} - \frac{1}{2}\left(\mathbf{I} - \hat{\boldsymbol{n}}_m\hat{\boldsymbol{n}}_m^{\mathrm{T}}\right). \tag{32}$$

Both the radial and angular integral are explicitly derived in appendix A2. The contribution from the autocorrelation of a single fiber, Eq. (28), becomes

$$\mathbf{N}_m = \zeta_m\frac{1}{2}\left(\hat{\boldsymbol{n}}_m\hat{\boldsymbol{n}}_m^{\mathrm{T}} - \frac{1}{3}\mathbf{I}\right). \tag{33}$$

As cylinder size is independent of orientation, $\zeta_m$ can be replaced by the mean volume fraction $\zeta_c$ when summing over *m* to yield the full mesoscopic contribution from autocorrelations

$$\mathbf{N}^{\mathrm{Auto}} = \zeta\frac{1}{2}\left(\mathbf{T} - \frac{1}{3}\mathbf{I}\right), \tag{34}$$

where $\mathbf{T}$ is the fODF scatter matrix[62]

$$\mathbf{T} = \left\langle \hat{\boldsymbol{n}}\hat{\boldsymbol{n}}^{\mathrm{T}} \right\rangle_{\hat{\boldsymbol{n}}} = \frac{1}{N}\sum_{m} \hat{\boldsymbol{n}}_m \hat{\boldsymbol{n}}_m^{\mathrm{T}}, \tag{35}$$

i.e., the second moment of the $N$ different orientations.

$\mathbf{N}^{\mathrm{Auto}}$ can also be rewritten in terms of the fODF $\mathcal{P}(\hat{\boldsymbol{n}})$ where it becomes

$$\mathbf{N}^{\mathrm{Auto}} = \zeta \frac{1}{2} \int d\hat{\boldsymbol{n}}\, \mathcal{P}(\hat{\boldsymbol{n}}) \left( \hat{\boldsymbol{n}}\hat{\boldsymbol{n}}^{\mathrm{T}} - \frac{1}{3}\mathbf{I} \right). \tag{36}$$

Using the Laplace expansion of $\mathcal{P}(\hat{\boldsymbol{n}})$,

$$\mathcal{P}(\hat{\boldsymbol{n}}) = \sum_{l=0,2,\ldots} \sum_{m=-l}^{l} \frac{(2l+1)}{4\pi} p_{lm}\, \mathrm{Y}_{lm}(\hat{\boldsymbol{n}}), \tag{37}$$

where $p_{lm}$ is the Laplace expansion coefficients,

$$\mathbf{N}^{\mathrm{Auto}} = \zeta \frac{1}{2} \sum_{l=0,2,\ldots} \sum_{m=-l}^{l} \frac{(2l+1)}{4\pi} p_{2m} \int d\hat{\boldsymbol{n}}\, \mathrm{Y}_{lm}(\hat{\boldsymbol{n}}) \left( \hat{\boldsymbol{n}}\hat{\boldsymbol{n}}^{\mathrm{T}} - \frac{1}{3}\mathbf{I} \right). \tag{38}$$

The product $\hat{\boldsymbol{n}}\hat{\boldsymbol{n}}^{\mathrm{T}}$ can be written in the basis of the irreducible rank-2 symmetric trace-free (STF) tensor representation of SO(3) $\boldsymbol{\mathcal{Y}}_{2m}$[63,64] defined by

$$\hat{\boldsymbol{n}}^{\mathrm{T}} \boldsymbol{\mathcal{Y}}_{2m} \hat{\boldsymbol{n}} = \mathrm{Y}_{2m}(\hat{\boldsymbol{n}}), \tag{39}$$

giving

$$\hat{\boldsymbol{n}}\hat{\boldsymbol{n}}^{\mathrm{T}} - \frac{1}{3}\mathbf{I} = \frac{8\pi}{15} \sum_{m=-2}^{2} \boldsymbol{\mathcal{Y}}_{2m} \mathrm{Y}_{2m}(\hat{\boldsymbol{n}}). \tag{40}$$

Plugging Eq. (38) into Eq. (36) yields

$$\mathbf{N}^{\mathrm{Auto}} = \zeta \frac{1}{3} \sum_{m=-2}^{2} p_{2m} \boldsymbol{\mathcal{Y}}_{2m}. \tag{41}$$

Hence, as $\mathbf{N}^{\text{Auto}}$ is a symmetric trace-free tensor of order $l = 2$, it depends only on the Laplace expansion coefficients $p_{2m}$ of the fODF $\mathcal{P}(\hat{\boldsymbol{n}})$.

*Mesoscopic contribution from cross-correlation*

Let us consider the cross-correlated mesoscopic contribution, Eq. (25), from two cylinders with distinct orientations. We present a qualitative reason for this correlation being zero already on the level of the structural correlation function (i.e. before convolution with the dipole kernel), i.e.

$$\Gamma_{mw}(\boldsymbol{r}) = \frac{1}{|\mathrm{M}|}\int d\mathbf{r}_0 v_m(\mathbf{r}_0) v_w(\mathbf{r}_0 + \mathbf{r}) = 0. \tag{42}$$

An explicit calculation with the same result is given in appendix A3.

To illustrate the present estimate, let us first consider the correlation of compact objects such as spheres or parallel cylinders considered in the two-dimensional volume of their cross-section. In this case, the value of $\int d\boldsymbol{r}_0\, v_m(\mathbf{r}_0) v_w(\mathbf{r}_0 + \mathbf{r})$ can have non-zero contributions at most for $\mathbf{r}_0$'s in a volume $\zeta V$. For each point in this volume , there is a probability on the order of $\zeta$ that $v(\mathbf{r}_0 + \mathbf{r})$ is also non-zero, which estimates the cross-correlation to be $\Gamma_{mw} \sim \zeta^2$.

Consider now the cross-correlation of nonparallel cylinders. As illustrated in Figure 3, non-zero contributions to $\int d\boldsymbol{r}_0\, v_m(\mathbf{r}_0) v_w(\mathbf{r}_0 + \mathbf{r})$ come only from small volumes at specific locations within the cylinders rather than their whole volume. This brings in the missing factor $\zeta^2 \rho / \mathrm{L}_\mathrm{M}$, where $\mathrm{L}_\mathrm{M}$ is the radius of the mesoscopic sphere. This ratio is zero in the limit of infinite cylinders.

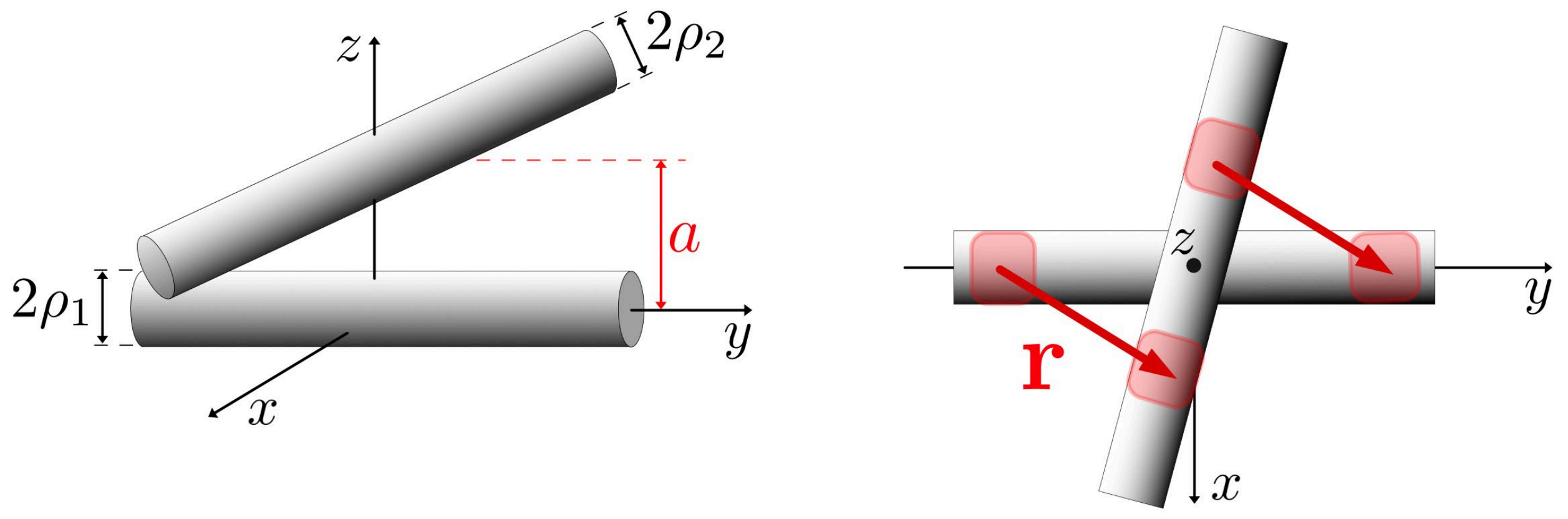

*Figure 3 - Contribution to the correlation function of two non-touching and nonparallel cylinders. The reference frame is selected in such a way that one cylinder (radius* $\rho_1$*) coincides with the y-axis while the other (radius* $\rho_2$*) is parallel to the x-y plane and shifted vertically by the distance* $a > \rho_1 + \rho_2$*. The right image shows the view from the tip of z-axis.* $\Gamma_{mw}(\mathbf{r})$ *is only nonzero for volumes in which* $r_z \approx a$ *(with a deviation about the cylinders' radii), for which it has contributions only for* $r_0$*'s at specific locations in the x-y plane. The contribution is thus not proportional to the cylinder volume.*

Note that parallel cylinders present a singular case in which the above argument for non-parallel cylinders does not work. To check and rule out the presence of a delta-functional contribution, we explicitly show in appendix A3 that the cross-correlated mesoscopic contribution, Eq. (25), from a collection of *randomly* positioned parallel cylinders is also zero. This ensures a smooth transition between randomly positioned parallel and non-parallel collections of cylinders, which is demonstrated in simulation (c). However, for a low number of cylinders, the transition would be singular as the total cross-correlations between the parallel cylinders would be non-zero, albeit second order in volume fraction $\Gamma_{mw} \sim \zeta^2$.

*In Total*

The mesoscopic contribution for a population of solid cylinders with orientation dispersion, randomly positioned, and with a diameter variation independent of their orientation is equal to the autocorrelation $\mathbf{N}^{\text{Auto}}$ part (Validated in simulation (c))

$$\mathbf{N} = \mathbf{N}^{\text{Auto}} = \zeta \frac{1}{3} \sum_{m=-2}^{2} p_{2m} \mathcal{Y}_{2m}. \tag{43}$$

The mesoscopic dipole coupling only depends on the *l=2* expansions coefficients of the fODF. $\mathbf{N}^{\text{Auto}}$ is zero only in the limit of maximal dispersion (i.e., randomly oriented cylinders). Hence, using the definition of the *l=2* STF tensors, Eq. (37), the mesoscopic NMR Larmor frequency $\bar{\Omega}^{\text{Meso}}(\mathbf{R})$, Eq. (12) , to the total Larmor frequency $\bar{\Omega}(\mathbf{R})$, Eq. (19), becomes

$$\bar{\Omega}^{\text{Meso}}(\mathbf{R}) = -\gamma B_0 \bar{\chi} \frac{1}{3} \sum_{m=-2}^{2} p_{2m}(\mathbf{R}) \mathrm{Y}_{2m}(\hat{\mathbf{B}}). \tag{44}$$

Equation (42) is also in agreement with previous results derived for parallel cylinders[45,60]. For a microstructure with axial symmetry around $\hat{z}$, only $p_{20}(\mathbf{R})$ is zero, which yields $\bar{\Omega}^{\text{Meso}}(\mathbf{R}) \propto p_{20}(\mathbf{R}) \mathrm{Y}_{20}(\hat{\mathbf{B}}) \propto \left(3\langle\cos^2(\theta)\rangle_{\hat{n}} - 1\right)\left(3\cos^2(\theta_{\hat{\mathbf{B}}}) - 1\right)$. Here $\theta_{\hat{\mathbf{B}}} = a\cos\left(\hat{z}\cdot\hat{\mathbf{B}}\right)$ is the polar angle between $\hat{z}$ and the external field, while $\langle\cos^2(\theta)\rangle_{\hat{n}}$ reflects the average polar angle of the cylinders to $\hat{z}$ [65]. This means $\bar{\Omega}^{\text{Meso}}(\mathbf{R})$ is zero at the magic angle $\theta_{\hat{\mathbf{B}}} \approx 54^\circ$, no matter the level of dispersion. For uniformly dispersed cylinders $p_{20}(\mathbf{R}) = 0$, while for fully parallel cylinders $p_{20}(\mathbf{R}) = 1$.

Combining the macroscopic contribution $\bar{\Omega}^{\text{Macro}}(\mathbf{R}) + \bar{\Omega}^{\text{W}}(\mathbf{R})$, Eq. (55), with the mesoscopic contribution $\bar{\Omega}^{\text{Meso}}(\mathbf{R})$, Eq. (42), the shift Eq. (19) in MRI Larmor frequency $\bar{\Omega}_{\text{MRI}}(\mathbf{R})$ finally becomes

$$\bar{\Omega}_{\text{MRI}}(\mathbf{R}) = \gamma \mathrm{B}_0 \left( -\bar{\chi}(\mathbf{R}) \frac{1}{3} \sum_{m=-2}^{2} p_{2m}(\mathbf{R}) \mathrm{Y}_{2m}(\hat{\mathbf{B}}) + \hat{\mathbf{B}}^{\mathrm{T}} \sum_{\mathbf{R}'} \bar{\Upsilon}(\mathbf{R}-\mathbf{R}') \bar{\chi}(\mathbf{R}') \hat{\mathbf{B}} \right) + \bar{\Omega}^{\text{W}}(\mathbf{R}). \tag{45}$$

Equation (43), revealing the dependence of the Larmor frequency on both susceptibility and local cylindrical microstructure, is our main theoretical result.

# Methods

**Simulations: Validation of modeling**

We designed 3 sets of simulations to validate our theory. We investigated **(a)** the influence of the size and shape of the mesoscopic Lorentz cavity $\mathrm{M}(\mathbf{R})$ when calculating the mesoscopic averaged Larmor frequency, Eq. (6) through Eqs. (10) and (11), for different microstructural configurations, **(b)** the accuracy of calculating the MRI Larmor frequency using Eq. (19) in a sample with slowly varying magnetic microstructure and run using different inclusion sizes compared to $\mathrm{M}(\mathbf{R})$. Lastly, we investigated in **(c)** the validity of the proposed mesoscopic kernel, Eq. (41), for a distribution of cylinders. All simulations were done in Matlab (The MathWorks, Natick, MA, USA).

*Random packing of Cylinders or spheres*

Non-overlapping cylinders or spheres were packed in a sample represented by a cubic volume of side length $\mathrm{L}$, with radii $\rho$ drawn from a gamma distribution with mean and standard deviation (SD) $\rho$. The positions $\boldsymbol{u}$ (parameterized as in appendix A for cylinders) were randomly generated from a uniform distribution, and the packing was stopped once the inclusions occupied a chosen volume fraction in the sample. For cylinders, various levels of orientation dispersion were achieved by only generating cylinders within an allowed polar-angle range $\cos^{-1}(\hat{\boldsymbol{z}}\cdot\hat{\boldsymbol{n}})=\theta<\theta_c$, set by a given cut-off $\theta_c$. This generated a uniform distribution of orientations up to $\theta_c$. The sample was subsequently discretized on a 3D grid, with a resolution set by the number of grid points. The indicator function $v(\boldsymbol{r})$ on the discretized grid was then Fourier transformed to yield the indicator function $v(\boldsymbol{k})$ in k-space, which in turn was used to compute the correlation function $\Gamma^{vv}(\boldsymbol{k})$, Eq. (17). We computed the local Larmor frequency $\Omega(\boldsymbol{r})=\gamma\hat{\mathbf{B}}^{\mathrm{T}}\Delta\mathbf{B}(\boldsymbol{r})$, Eq. (1), by multiplying $v(\boldsymbol{k})$ with the dicretized dipole kernel $\hat{\mathbf{B}}^{\mathrm{T}}\boldsymbol{\Upsilon}(\boldsymbol{k})\hat{\mathbf{B}}$, Eq. (3), and Fourier transforming back to image space. The mesoscopic dipole tensor $\mathbf{N}$, Eq. (16), was computed by integrating the product of the dipole kernel and the correlation function, and dividing the result by the volume fraction $1-\zeta$ of the discretized grid. The external field is normalized so $\gamma\mathrm{B}_0=1\dfrac{\mathrm{rad}}{\mathrm{s}}$, and chosen to be along $\hat{\mathbf{B}}=\hat{\boldsymbol{x}}$ to maximize effect size.

*Simulation (a) The size of* $\mathrm{M}(\mathbf{R})$

The purpose of this simulation was to examine the accuracy of the mesoscopic Lorentz cavity approximation, Eq. (6) through Eqs. (10) and (11) as a function of the size of $\mathrm{M}(\mathbf{R})$ compared to the size of the magnetic inclusions. Figure 4 gives an overview of the simulation. The coarse-grained Larmor frequency $\bar{\Omega}(\mathbf{R})$ was averaged around the center point $\mathbf{R}$ of a population of (**a1**) randomly placed parallel rectangular rods, (**a2**) randomly placed dots, (**a3**) long randomly oriented rods and (**a4**) random sheets parallel along the $\hat{z}$ direction. The rods (**a1**) were made by generating a $400^2$ 2D grid with a weighted random number of zeros and ones to vary the volume fraction from $\zeta \approx 0.1,...,0.9$. The 2D grid was then repeated along the third dimension to yield a $400^3$ grid with purely 2D Poissonian structural correlations[46]. The random dots (**a2**) were generated as (**a1**) but placed randomly on a 3D $400^3$ grid with similar volume fractions. Each rod/dot thus had a width $2\rho$ of 1 grid unit and the rods a length of 400 grid units. The long randomly oriented rods (**a3**) were made by randomly placing rods with a width of 3 grid units and length of 800 grid units in a $400^3$ 3D grid. The rods were allowed to overlap, and we used periodic boundaries to contain the whole rod. The volume fraction was varied from $\zeta \approx 0.1,...,0.9$. The sheets (**a4**) were made by placing rods of similar dimensions in a $400^2$ 2D grid and repeating it along the third dimension to yield a $400^3$ grid with similar volume fractions. These scenarios correspond to systems with long-range 3D and 2D structural correlations, respectively. 100 different configurations were generated for (**a1)-(a4**) for each volume fraction. When calculating $\bar{\Omega}(\mathbf{R})$, the microstructure $v(\boldsymbol{r})$ was only considered within the mesoscopic region $\mathrm{M}(\mathbf{R})$ (Figure 4A). The exterior region was instead averaged according to Eq. (8). We investigated the dependence of the size and shape of $\mathrm{M}(\mathbf{R})$ when calculating $\bar{\Omega}(\mathbf{R})$ in comparison to the true Larmor frequency $\bar{\Omega}_{\mathrm{True}}(\mathbf{R})$ calculated with Eq. (6) using the magnetic microstructure for the entire region and averaged within the same mesoscopic sphere $\mathrm{M}(\mathbf{R})$ as $\bar{\Omega}(\mathbf{R})$ (Figure 4B). In addition, we investigated the influence of various shapes for $\mathrm{M}(\mathbf{R})$, i.e. a cube, sphere and cylinder, for decomposing the sample into mesoscopic and macroscopic regions in the calculation of $\bar{\Omega}(\mathbf{R})$. The side length $\mathrm{L}_{\mathrm{M}}$ of $\mathrm{M}(\mathbf{R})$ was varied from 4 times the width of the rods to a half the size of the sample. We also computed the field at the center point $\mathbf{R}$, $\Omega_{\mathrm{Center}}(\mathbf{R})$ where the microstructure was removed within $\mathrm{M}(\mathbf{R})$ (Figure 4C). The mean relative error $\left\langle\left(\left(\bar{\Omega}^{\text{Meso}}+\bar{\Omega}^{\text{Macro}}\right)/\bar{\Omega}_{\text{True}}\right)\right\rangle$ (for **(a1)**) were then calculated averaging over the different realizations for each configuration. We also computed the mean squared difference $\left\langle\varepsilon^2\right\rangle=\left\langle\left(\left(\bar{\Omega}^{\text{Meso}}+\bar{\Omega}^{\text{Macro}}-\bar{\Omega}_{\text{True}}\right)^2\right)\right\rangle$ and compared it to $\left\langle\Omega^2_{\text{Center}}\right\rangle$ (for **(a1)-(a4)**).

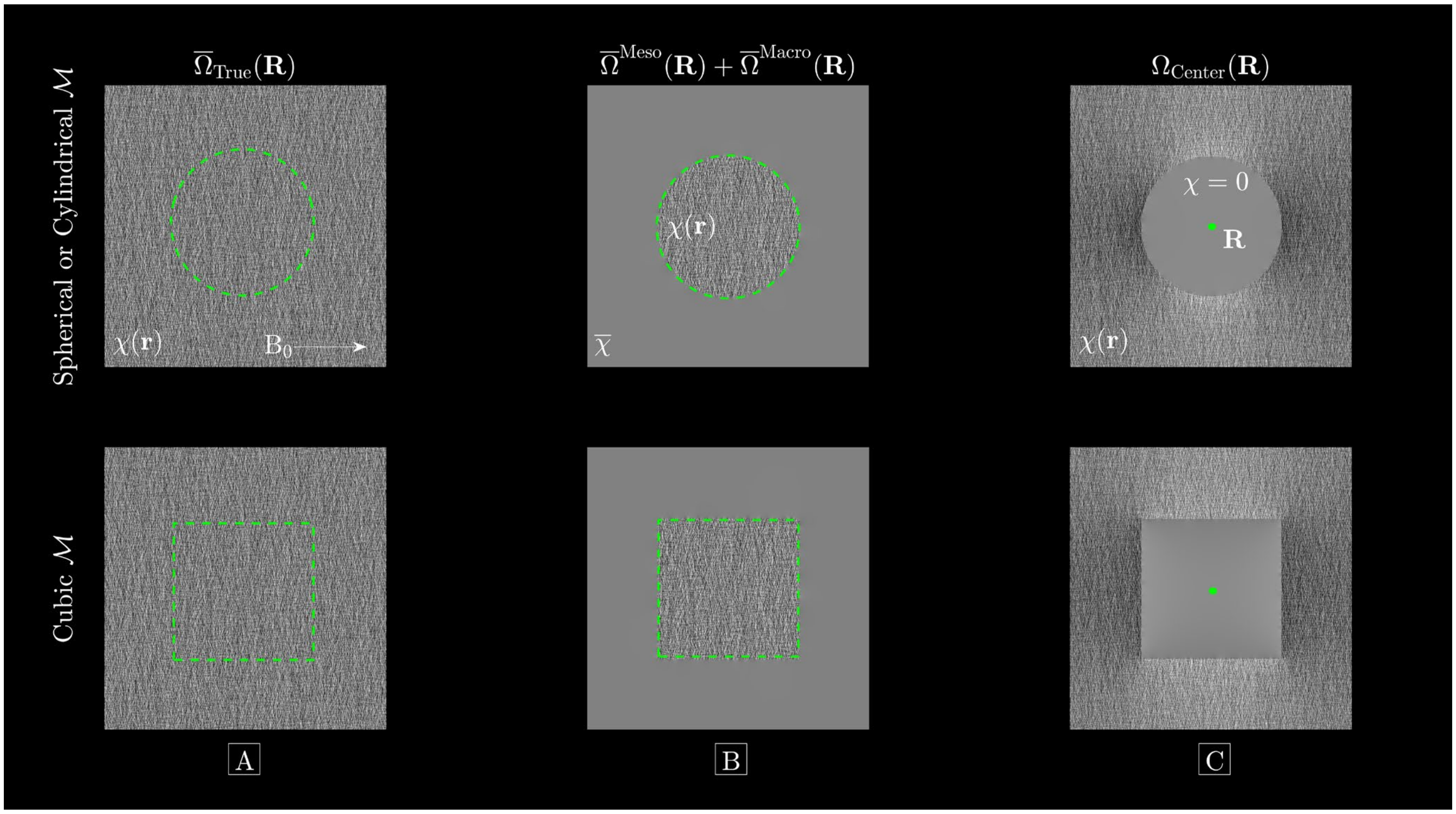

*Figure 4 - Simulation (a). Images show a cross-section of the Larmor frequency for a Poissonian system of parallel rods or random dots used to test the influence of the size of* $\mathrm{M}(\mathbf{R})$*, here given by the green circle or box.* ***A*** *shows the frequency by averaging within* $\mathrm{M}(\mathbf{R})$ *including the whole magnetic microstructure. In* ***B****, the volume outside* $\mathrm{M}(\mathbf{R})$ *has been coarse grained. In* ***C****, the magnetic microstructure has been nulled within* $\mathrm{M}(\mathbf{R})$*, and the frequency is calculated at the point* $\mathbf{R}$*. The frequency is induced by an external field* $\mathbf{B}_0=\mathrm{B}_0\hat{\boldsymbol{x}}$*. The mesoscopic region is tested both as a cube, sphere, and cylinder.*

*Simulation (b) The voxel averaged Larmor frequency in a sample with varying microstructure*

The purpose of the next simulation was to validate Eq. (19) describing the MRI measured Larmor frequency when the sample consists of a slowly varying magnetic microstructure, and quantify the error we obtain when (**b1**) we filter the measured Larmor frequency (as described in Eq. (6)); (**b2**) neglect the sampling PSF (as explained in appendix A2) and lastly; (**b3**) partition the magnetic microstructure in order to compute the macroscopic contribution (cf. Eq. (55) in appendix A2) to the measured Larmor frequency.

Figure 5 illustrates the sample used to test the error from these three points. We constructed a sample consisting of 27 cubes, denoted $\tilde{\mathrm{U}}_q$, stacked $3\times3\times3$ (blue square in Figure 5A). Each cube had side length $3\mathrm{L}$ (see Figure 5) and contained a configuration of cylinders with a randomly chosen axially symmetric orientation dispersion and volume fraction $\zeta\approx0.15$ (i.e., different for each cube). Susceptibility variations were introduced by randomly choosing a normalized microscopic susceptibility $\chi\in(-1;1)$ for each cube (in reality $\chi\sim\mathrm{ppm}$, but as we only consider field ratios, the absolute magnitude is irrelevant). The sample was further partitioned into smaller cubes $\mathrm{U}_q$ with center points denoted $\mathbf{R}_q$ and with side length $\mathrm{L}$, i.e. one third of the larger cubes $\tilde{\mathrm{U}}_q$ (indicated by the orange square in Figure 5A). This construction ensured a sample with a more slowly varying microstructure across $\mathrm{U}_q$, as seen at $\mathbf{R}_j$, since the abrupt edges between the cubes $\mathrm{U}_q$ are far away. Thus, every large cube $\tilde{\mathrm{U}}_q$ with a distinct magnetic microstructure contained 27 smaller cubes $\mathrm{U}_q$, and so the entire sample was partitioned into 729 cubes $\mathrm{U}_q$ stacked $9\times9\times9$.

As illustrated in Figure 5A, the aim was to calculate the MRI Larmor frequency $\bar{\Omega}_{\mathrm{MRI}}\left(\mathbf{R}_j\right)$ at the center at $\mathbf{R}_j$ in the sample, defined by a PSF assumed here to be a 3D sinc as would be obtained from finite and discrete Cartesian k-space sampling, and with volume equal to $\mathrm{U}_j$. Calculating the MRI Larmor frequency using Eq. (54) from appendix A2, including the PSF and the full "microscopic" magnetic microstructure in the sample defined the ground truth frequency $\bar{\Omega}_{\mathrm{MRI}}\left(\mathbf{R}_j\right)$.

As specified in Figure 5B, we tested our model for the estimated Larmor frequency, Eq. (19), in the following way: The mesoscopic contribution $\bar{\Omega}^{\mathrm{Meso}}\left(\mathbf{R}_j\right)$ was calculated using Eq. (18) as a circular convolution between the discretized magnetic microstructure $\chi(\boldsymbol{r}')$ from the cube $\mathrm{U}_j$ and dipole kernel. This meant that the size of $\mathrm{M}$ averaging the Larmor frequency (**b1**), and the mesoscopic cavity in which explicit magnetic microstructure was used to compute $\bar{\Omega}^{\text{Meso}}\left(\mathbf{R}_j\right)$, effectively was the same size as $\mathrm{U}_j$. As shown in Figure 5C, the macroscopic contribution $\bar{\Omega}^{\text{Macro}}\left(\mathbf{R}_j\right)$ was calculated from the sum in Eq. (55) using bulk susceptibilities $\bar{\chi}\left(\mathbf{R}_q\right)$ Eq. (8), estimated by coarse-graining the magnetic microstructure with $\mathrm{M}$ chosen to be the same size as $\mathrm{U}_q$. This then neglected the PSF (**b2**) and partitioned the magnetic microstructure (**b3)** when computing $\bar{\Omega}^{\text{Macro}}\left(\mathbf{R}_j\right)$. To highlight the importance of using the voxel averaged dipole field, Eq. (56), we calculated $\bar{\Omega}^{\text{Macro}}\left(\mathbf{R}_j\right)$ using either the averaged or the elementary dipole field, Eq. (2). Last, we also considered neglecting the mesoscopic contribution, which resembled a QSM-like calculation. To underscore the sensitivity of the approximation to the size of $\mathrm{U}_q$ relative to the size of the cylinders, we performed the calculation for various ratios $\mathrm{L}/\rho$ ranging from 3 to 40. The simulations were performed 10 times for each ratio to calculate the mean ratio between the model and ground truth, $\left\langle\left(\left(\bar{\Omega}^{\text{Meso}}\left(\mathbf{R}_j\right)+\bar{\Omega}^{\text{Macro}}\left(\mathbf{R}_j\right)\right)\big/\bar{\Omega}_{\text{MRI}}\left(\mathbf{R}_j\right)\right)\right\rangle$ along with the SD. The sample was discretized with a fixed resolution with respect to $\mathrm{L}$. The resolution thus ranged from $0.33\rho$ to $0.025\rho$ as $\mathrm{L}/\rho$ ranged from approximately 3 to 35, respectively.

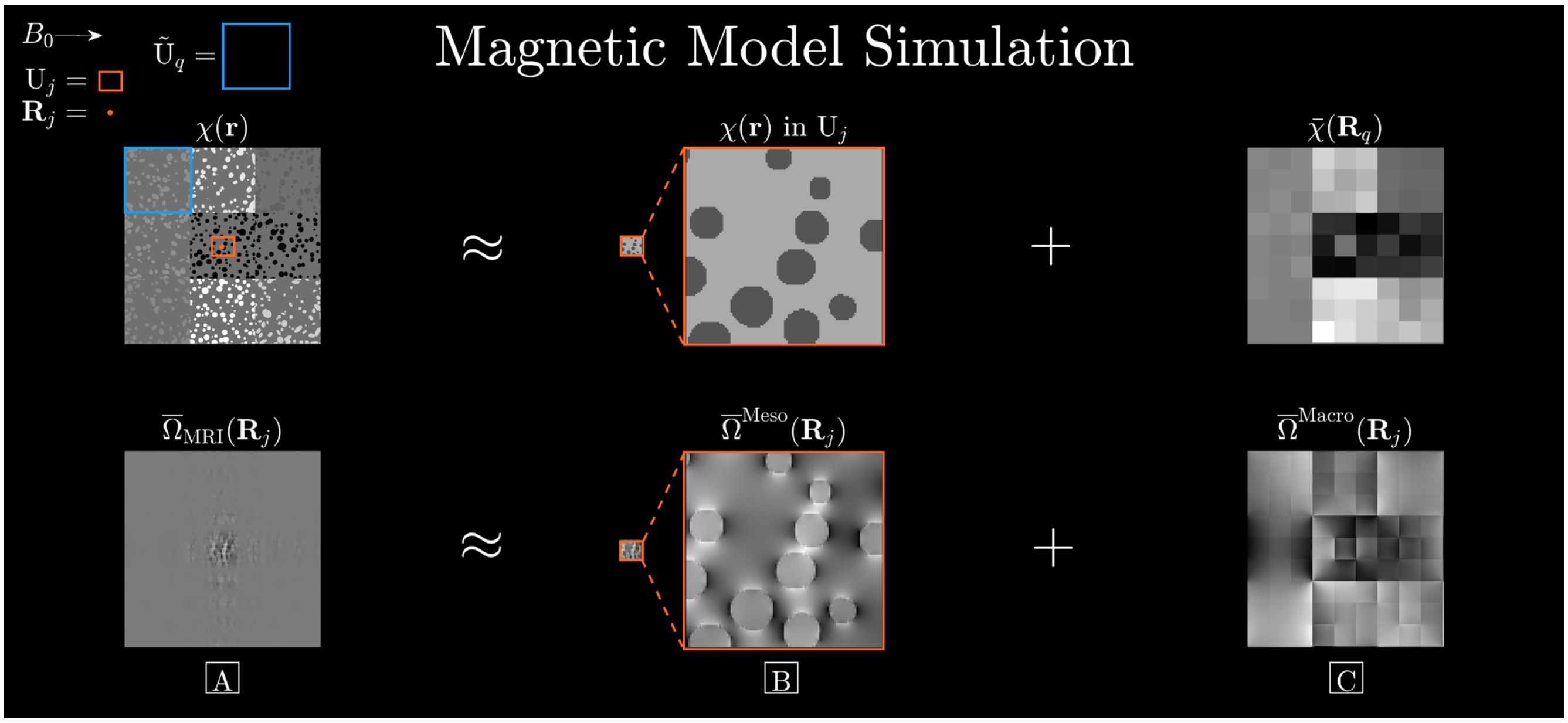

*Figure 5 - Simulation (b). Overview of Magnetic Model Simulation. In **A**, a cross section of the microstructure* $\chi(\boldsymbol{r})$ *can be seen. Here 9 different regions* $\tilde{\mathrm{U}}_q$ *(blue square) with random susceptibility and orientation dispersion are clearly visible. We are interested in calculating the average Larmor frequency sampled by the PSF* $\mathrm{P}(\boldsymbol{r}-\mathbf{R}_j)$ *positioned at the center point denoted* $\mathbf{R}_j$. $\bar{\Omega}_{\mathrm{MRI}}(\mathbf{R}_j)$ *is obtained by the induced field multiplied by the PSF:* $(1-v(\boldsymbol{r}))\Omega(\boldsymbol{r})\mathrm{P}(\boldsymbol{r}-\mathbf{R}_j)$*, whose average gives* $\bar{\Omega}_{\mathrm{MRI}}(\mathbf{R}_j)$*. In **B**, the microstructure and Larmor frequency within the center region* $\mathrm{U}_j$ *is shown, defining the mesoscopic contribution* $\bar{\Omega}^{\mathrm{Meso}}(\mathbf{R}_j)$ *to the average field.* $\mathrm{U}_j$ *has been magnified to highlight the microstructure. In **C**, the macroscopically averaged microstructure is shown along with the associated Larmor frequency* $\bar{\Omega}^{\mathrm{Macro}}(\mathbf{R}_j)$*. This is used to calculate the frequency contribution from neighboring cubes* $\mathrm{U}_q$ *at the point* $\mathbf{R}_j$*, indicated by the orange dot.*

*Simulation (c) Validity of the proposed mesoscopic kernel for a distribution of cylinders*

In the final simulation, we considered 24 populations of cylinders with different levels of orientation dispersion to test our results for the proposed mesoscopic kernel $\mathbf{N}/\zeta$, Eq. (41), compared to Eq. (16), which takes into account the full correlation function, Eq. (17). Figure 6 gives an overview of the different populations of cylinders. Dispersion was varied from fully parallel to maximal orientation dispersion (isotropic). The volume fraction was $\zeta \approx 0.15$. Every configuration was discretized with a resolution of $0.08\rho$ . The eigenvalues $\left(\lambda_{\perp 1}, \lambda_{\perp 2}, \lambda_{\parallel}\right)$ of $\mathbf{N}/\zeta$ using Eq. (41) or Eq. (16) were then calculated for each population and compared.

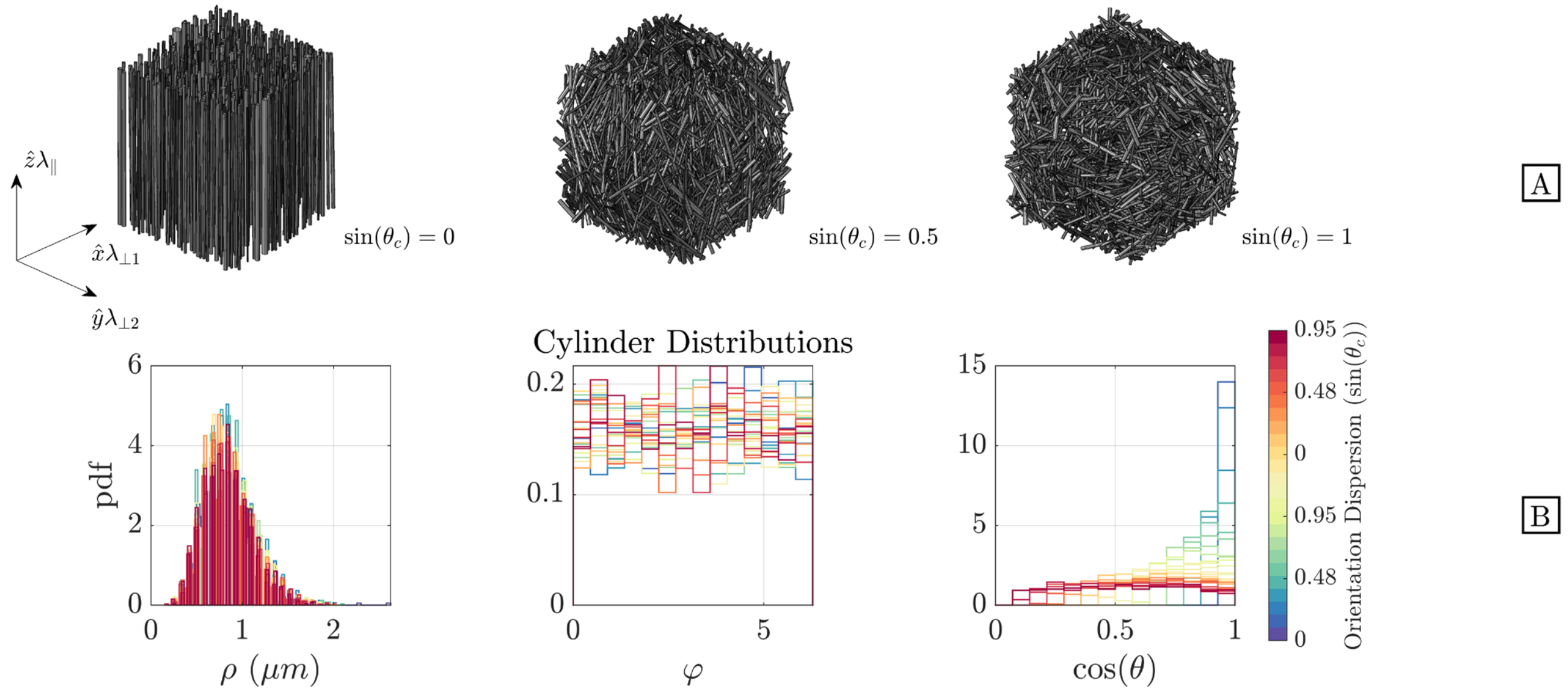

*Figure 6 - Simulation (c). Populations of cylinders with different levels of orientation dispersion are shown in* ***A***. *In* ***B***, *probability density function (pdf) of the resulting cylinder parameters for each configuration are shown. The cylinder radius $\rho$ is gamma-distributed, while $\varphi$ and $\theta$ are uniformly distribution in the full range of azimuthal angle and from zero to the maximum polar angle $\theta_c$, respectively. Colors are used to represent different populations with orientation dispersion indicated by the color bar.*

# Results

## Simulations

### *Simulation (a) The size of* $\mathbf{M}(\mathbf{R})$

Figure 7 shows the ratio of model and ground truth Larmor frequency for different sizes of the mesoscopic region and volume fraction for a bundle of uniformly positioned parallel rectangular rods. For a small mesoscopic cavity, the relative error from replacing the nearby microstructure with a homogenous medium with the same average susceptibility is high. As the mesoscopic cavity increases, the actual positions of the cylinders become less important, and the ratio can be seen to converge to 1. The SD is clearly lowest for high volume fraction due to a larger number of involved rods. Additionally, it is clear the coarse grained Larmor frequency can be decomposed using any cavity shape for $\mathrm{M}(\mathbf{R})$.

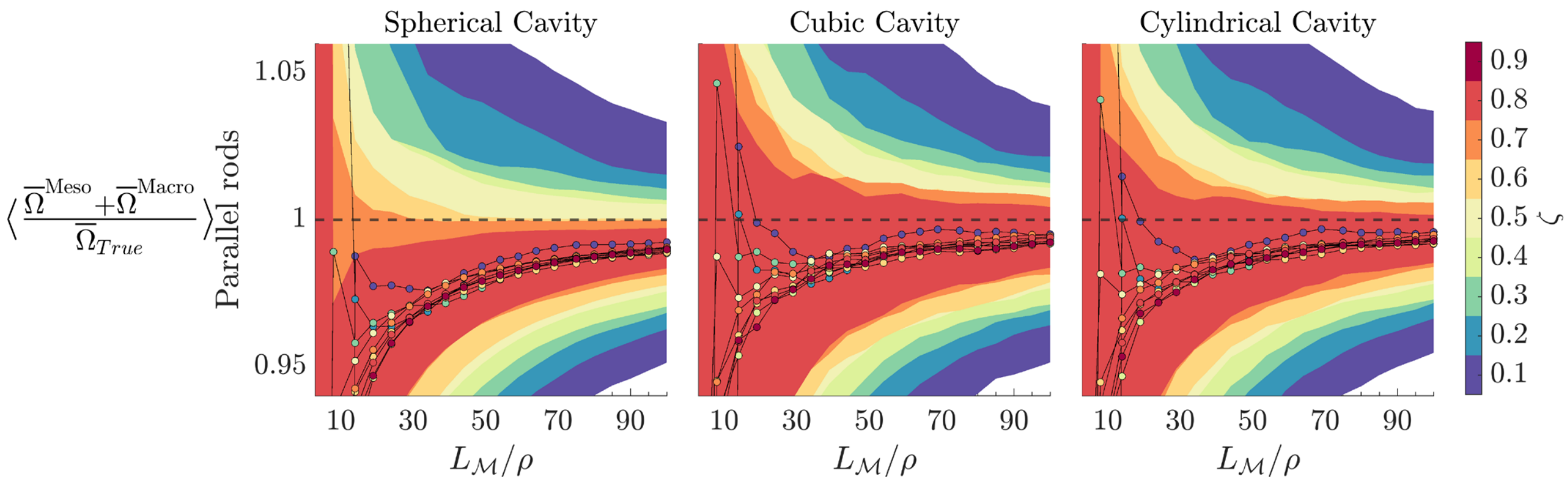

*Figure 7 - Simulation (a). The mean error (dots) and standard deviation (colorbands) of the estimated Larmor frequency is shown as a function of the size of the mesoscopic region* $\mathrm{L_M}$ *compared to the width of the rods* $2\rho$. *The error can be seen to converge to below 5% when* $\mathrm{L_M}$ *is around 10 times* $\rho$ *for all three cavity geometries. This underscores that the shape of the mesoscopic region is irrelevant, only size matters.*

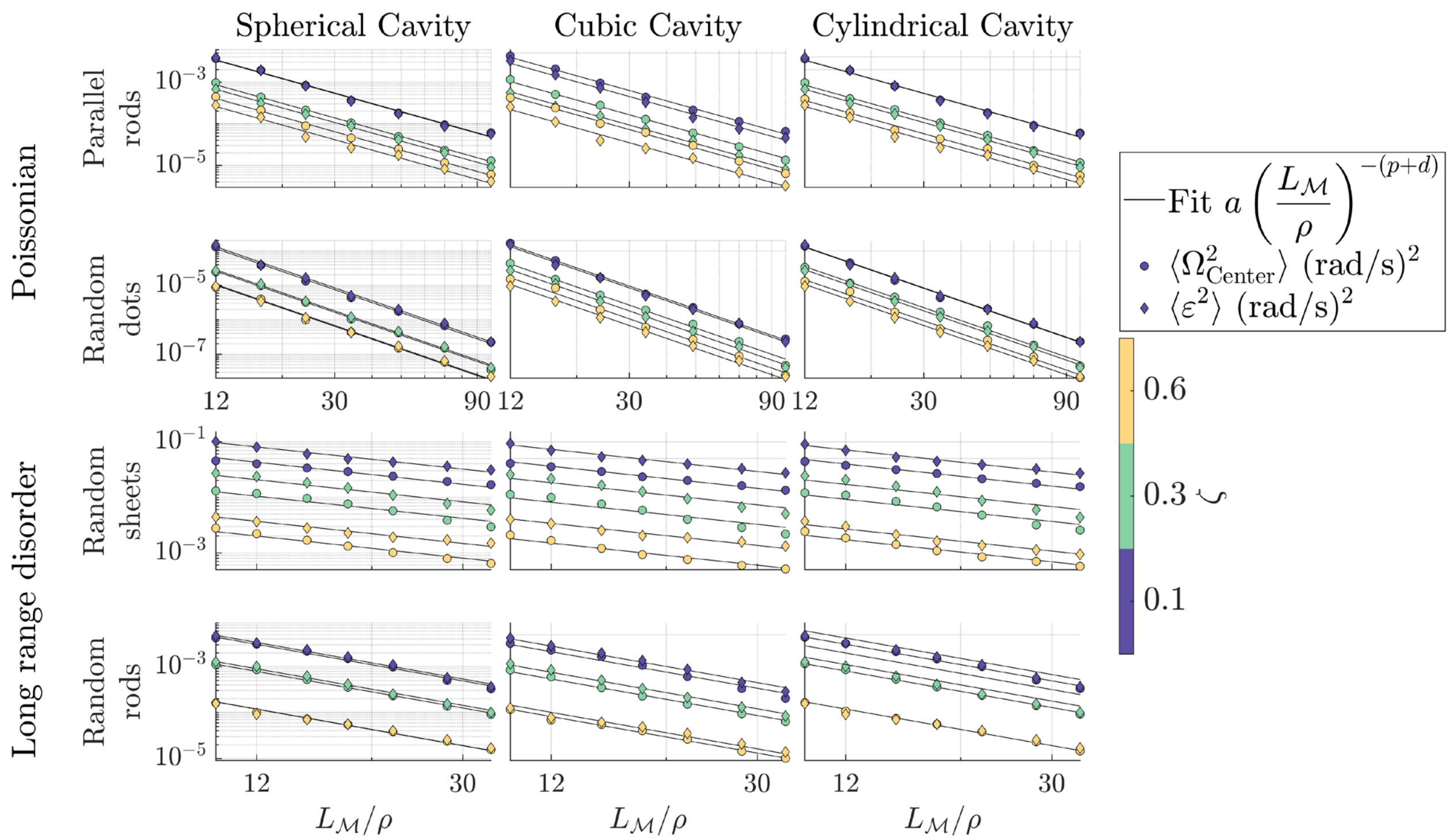

*Figure 8 - simulation (a). The y-axis corresponds to the mean squared difference* $\left\langle \varepsilon^2 \right\rangle = \left\langle \left( \left( \bar{\Omega}^{\text{Meso}} + \bar{\Omega}^{\text{Macro}} - \bar{\Omega}_{\text{True}} \right)^2 \right) \right\rangle$*, the center field* $\left\langle \Omega^2_{\text{Center}} \right\rangle$ *without microstructure within mesoscopic cavity* $\mathbf{M}(\mathbf{R})$*. The black lines are from fitting* $a\left(\frac{\mathrm{L_M}}{\rho}\right)^{-(p+d)}$ *to* $\left\langle \varepsilon^2 \right\rangle$ *and* $\left\langle \Omega^2_{\text{Center}} \right\rangle$*. Hence, only a single parameter is fitted, while the scaling is predicted by the structural universality class* $p+d$*, for* $d$ *dimensions and with* $p$ *being the structural exponent of the microstructure*[46]*. The x-axis denotes the size of the mesoscopic region* $\mathrm{L_M}$ *compared to the width of the rods* $2\rho$*. The first two rows show 2D (parallel rods) and 3D (random dots) Poissonian disorder. The last two rows show long range disorder, where 2D corresponds to sheets and 3D is long randomly oriented sticks. Three different volume fractions are shown for better visualization.*

Figure 8 shows that the mean squared difference $\left\langle \varepsilon^2 \right\rangle$ (cf. Figure 4A-B) and mean center field $\left\langle \Omega^2_{\text{Center}} \right\rangle$ (cf. Figure 4C) converge towards zero, no matter the cavity shape and dimensionality, with the convergence rate corresponding to the sum $p+d$, i.e. $\left\langle \varepsilon^2 \right\rangle \sim \left\langle \Omega^2_{\text{Center}} \right\rangle \sim \left(\frac{\mathrm{L_M}}{\rho}\right)^{-(p+d)}$. Here, $p$ is the structural disorder exponent and $d$ the dimensionality[46]. In appendix C we derive this scaling relation.

For the parallel rods simulation, the dimensionality is $d = 2$ with $p = 0$ or, equivalently $d = 3$ and $p = -1$, while for the dots, $d = 3$ and $p = 0$. For the randomly oriented rods and sheets in $d = 3$, $p = -1$ and $p = -2$, respectively.

*Simulation (b) The voxel averaged Larmor frequency in a sample with varying microstructure*

Figure 9 shows the mean ratio $\left\langle\left(\left(\bar{\Omega}^{\text{Meso}}\left(\mathbf{R}_j\right)+\bar{\Omega}^{\text{Macro}}\left(\mathbf{R}_j\right)\right)\middle/\bar{\Omega}_{\text{MRI}}\left(\mathbf{R}_j\right)\right)\right\rangle$, along with SD, for different ratios $\mathrm{L}/\rho$ when estimating the MRI Larmor frequency through a mesoscopic (Eq. (18)) and macroscopic (Eq. (55)) contribution. This was done based on 10 samples for each $\mathrm{L}/\rho$. The green curve shows the full model with voxel averaged dipole field, Eq. (19). For these particular samples, our model under- or overestimated the ground truth field with a SD less than around 3-8% once the voxel size was around 31-35 times the average radius of the cylinders. The red curves shows the full model but with the elementary field Eq. (2), for calculating the macroscopic contribution. We found it overestimated the MRI Larmor frequency, around 20-30% with SD around 40-50%. Blue shows neglecting both the mesoscopic contribution and using the elementary field, and hence resembles QSM calculation. It equally over and underestimates the Larmor frequency around 25% both with a SD up to 100%.

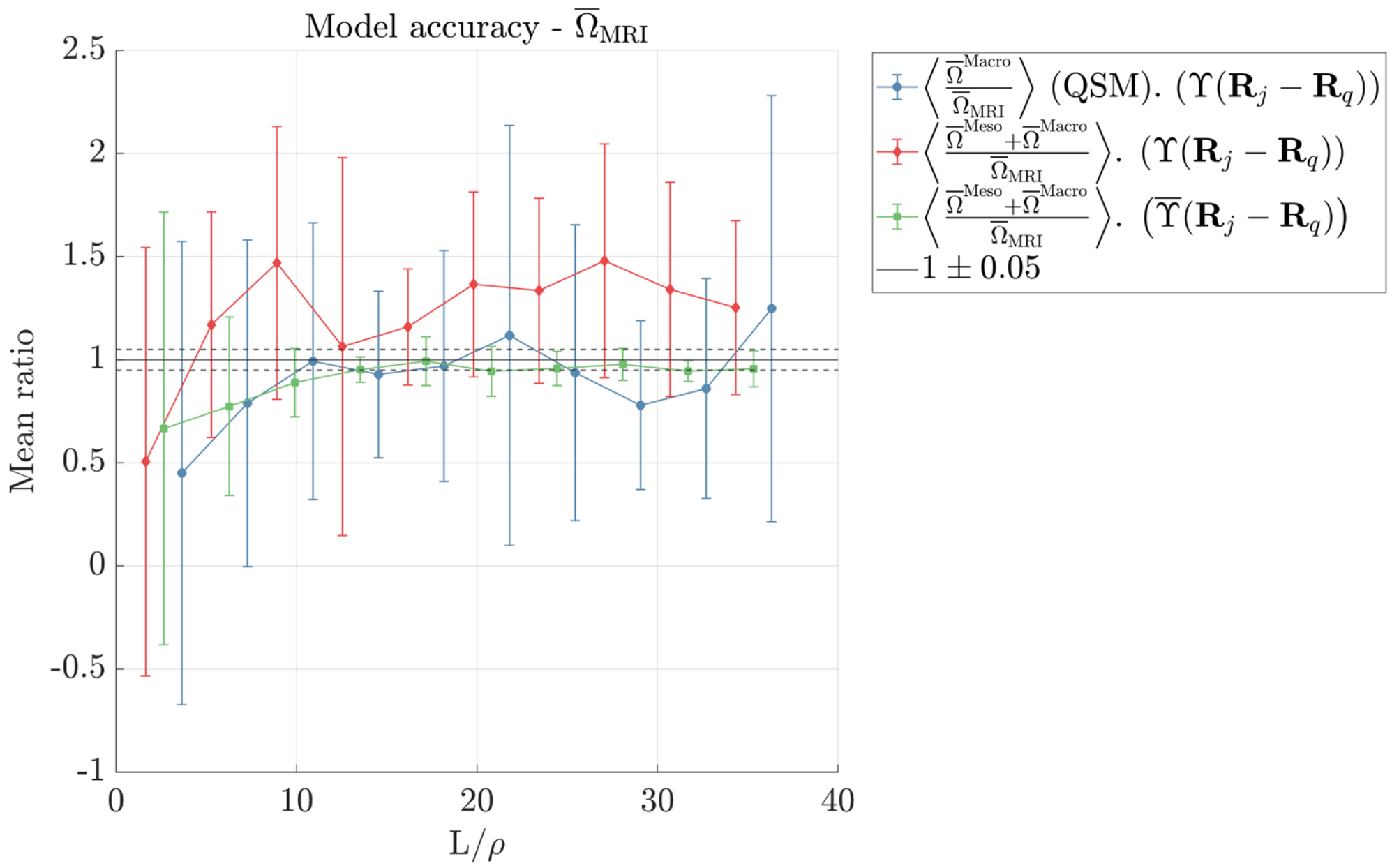

*Figure 9 - Simulation (b). Mean ratio (circles) and SD (vertical lines) on the average Larmor frequency estimate in the center voxel of a sample with smoothly varying microstructure (see Figure 5). The red and blue curve are slightly shifted to the left and right for better visualization. The average field within the center voxel is calculated using the proposed model* $\bar{\Omega}^{\mathrm{Meso}}\left(\mathbf{R}_j\right)+\bar{\Omega}^{\mathrm{Macro}}\left(\mathbf{R}_j\right)$*, and compared with the ground truth calculation* $\bar{\Omega}_{\mathrm{MRI}}\left(\mathbf{R}_j\right)$ *to compute their ratio. 10 samples were used for each* $L/\rho$ *(sampling resolution/inclusions size). Black indicates using the full model with voxel averaged dipole field, Eq. (56). Red indicates using the full model with the elementary field Eq. (2), for calculating the macroscopic contribution* $\bar{\Omega}^{\mathrm{Macro}}\left(\mathbf{R}_j\right)$*, Eq.(55). Blue indicates neglecting both the mesoscopic contribution and using the elementary field.*

*Simulation (c) Validity of the proposed mesoscopic kernel for a distribution of cylinders*

Figure 10 shows the eigenvalues $\left(\lambda_{\perp 1}, \lambda_{\perp 2}, \lambda_{\parallel}\right)$ of $\mathbf{N}/\zeta$ from the 24 simulations of cylinders with different orientation dispersion. It is clearly seen that the model follows the same behavior as the ground truth. A slight systematic discrepancy in the eigenvalues can be seen for low dispersion $\left(0<\theta_c<0.3\right)$. However, the maximum discrepancy was less than 4% and can be explained by the truncation of the cylinders at the edge of the finite sample space, as the magnitude of the eigenvalues is smaller for a finite cylinder compared to infinite cylinders[58,61]. As the orientation dispersion is increased, this effect is largely cancelled due to conservation of trace[52], i.e. $\mathrm{Tr}[\mathbf{N}/\zeta]=0$. The principal eigenvector of $\mathbf{N}/\zeta$ also agreed with ground truth, with an angular error being less than 2.5 degrees on average (data not shown).

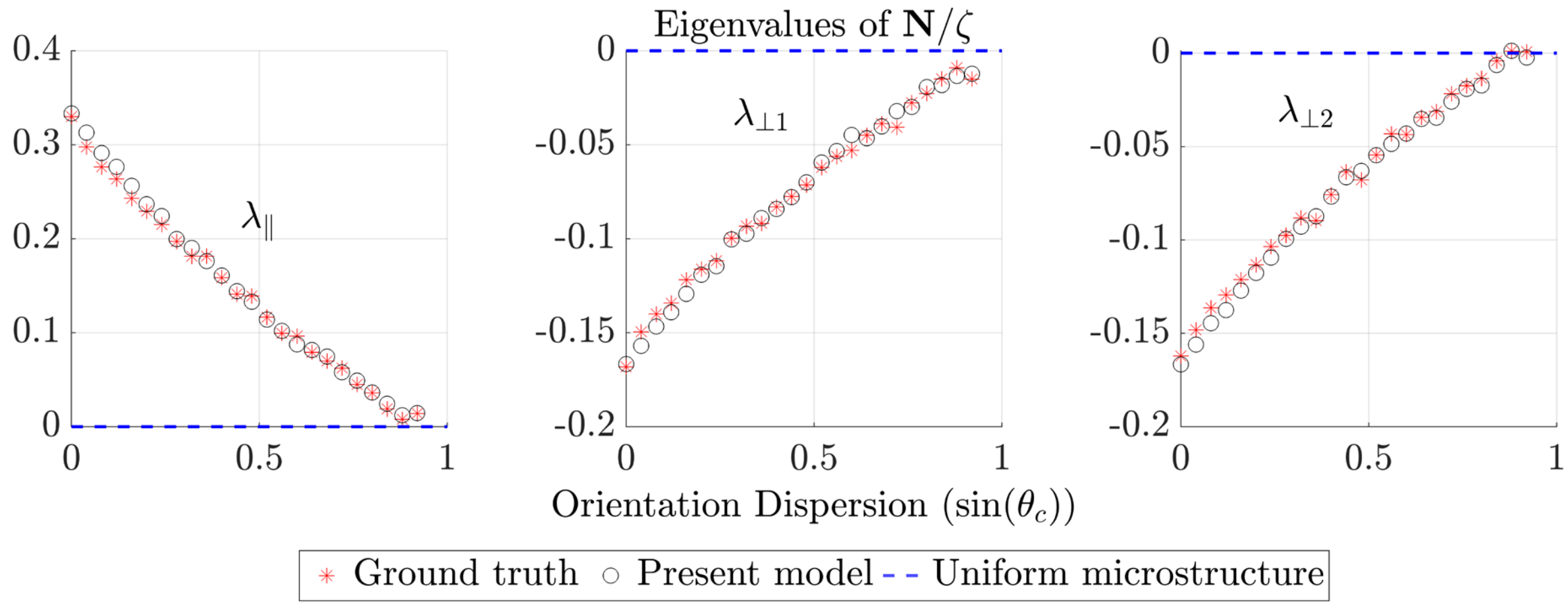

*Figure 10 - Simulation (c). Simulation of the mesoscopic contribution from 24 different orientation distributions. Eigenvalues of* $\mathbf{N}/\zeta$ *are presented for various levels of dispersion* $\sin(\theta_c)$ *set by the maximum allowed polar angle* $\theta_c$. *Notice the trace of* $\mathbf{N}/\zeta$ is zero, as expected from Eq. (32).

## Discussion

### Theory

We considered the local microscopic Larmor frequency $\Omega(\boldsymbol{r})$ at allowed spin positions outside magnetized inclusions, i.e., given by the "complement" $1-v(\mathbf{r})$ of the indicator function $v(\mathbf{r})$. The motivation was to reiterate how the MR measurable Larmor frequencies are affected by $\Omega(\boldsymbol{r})$ induced by magnetized tissue, here using the principle of coarse graining. We employ the magnetostatic formalism of working directly with magnetization (cf. Eq. (1)), instead of characterizing the magnetization in terms of volume and surface magnetic charge contributions. For an in-depth explanation of the different magnetostatic formalisms, we refer to Jackson Electrodynamics[66] 3rd Edition.
The MR signal was considered for weakly magnetized tissue with a convergent cumulant expansion; a requirement which is fullfilled for short times regardless of other parameters. It is also valid for any time when spins diffuse long distances across the entire range of frequencies acquiring a small phase over the correlation length $l_c$ of the medium (the motional- or diffusion narrowing regime[55] (DNR)). In this case, the first term of the cumulant expansion for the signal coincides with the first term of the signal expansion in the static dephasing regime (SDR), namely the sample (or voxel) averaged Larmor frequency.
To estimate the higher-order terms in the diffusion narrowing regime, consider a generic random medium, where the scaling of the n'th cumulant[55] is on the order of $\langle \phi \rangle^n \sim \delta\Omega^n \tau_c^{n-1} t$. If we assume a typical frequency $\delta\Omega \sim 0.3\,\mathrm{rad/ms}$ and a correlation time $\tau_c \sim l_c^2 / D \sim 1\,\mathrm{ms}$ (which are typical values[31,67] on a clinical scanner for myelinated axons at room temperature), then the ratio of the third to the first cumulant is $\delta\Omega^2 \tau_c^2 \sim 0.09$, i.e. on the order of 10% indicating a reasonable applicability of DNR. Reaching SDR can be achieved if $t \ll \tau_c$, as the scaling of the n'th cumulant becomes $\langle \phi \rangle^n \sim \delta\Omega^n t^n$. Such sub-millisecond echo times may be infeasible to achieve using conventional MRI acquisitions, instead, alternative ultra-fast acquisitions schemes reaching echo times on the order of microseconds may be considered[68].
As the signal is measured on the macroscale, the first cumulant can be described as an average of mesoscopically averaged microscopic frequencies, $\bar{\Omega}(\mathbf{R})$. We considered the case when the magnetic microstructure is statistically homogenous with scalar susceptibility $\chi(\boldsymbol{r}) = \chi v(\boldsymbol{r})$, and slowly varying across the macroscopic sample. We decomposed $\bar{\Omega}(\mathbf{R})$ into two contributions: The first described the mesoscopic contribution $\bar{\Omega}^{\mathrm{Meso}}(\mathbf{R})$, Eq. (12), depending explicitly on the positions and susceptibility of inclusion within a mesoscopic sphere positioned at $\mathbf{R}$, i.e. the magnetic microstructure. We show that $\bar{\Omega}^{\mathrm{Meso}}(\mathbf{R})$ depends on the magnetic susceptibility $\chi$, and a mesoscopic demagnetization tensor $\mathbf{N}$ (Eq. (16)) capturing the dependence on microstructure, in agreement with previous studies[38,45,60]. This contribution is of importance as studies have shown magnetic microstructure effects contribute substantially to the Larmor frequency in WM[40]. Additionally, the macroscopic contribution $\bar{\Omega}^{\mathrm{Macro}}(\mathbf{R})$

in Eq. (10) describing the Larmor frequency shift induced by inclusions outside the mesoscopic sphere was described by a convolution between the coarse grained magnetic microstructure $\bar{\chi}(\mathbf{R})$ and the dipole field $\Upsilon(\mathbf{R}-\mathbf{R}')$.

If we compare the mesoscopic Larmor frequency $\bar{\Omega}(\mathbf{R})$, Eq. (6), to the microscopic Larmor frequency $\Omega(\boldsymbol{r})$, Eq. (1), they both depend on a convolution between the elementary dipole field and magnetic susceptibility coarse-grained to the corresponding length scale (i.e., either $\bar{\chi}(\mathbf{R})$ or $\chi(\boldsymbol{r})$, respectively). In the microscopic relation, $\Omega(\boldsymbol{r})$, a "molecular contribution" is in principle present[50] , but as we are concerned with isotropic liquids, this contribution is zero. Likewise, for $\bar{\Omega}(\mathbf{R})$, we obtain a microstructural contribution, Eq. (11), which is only zero for uniform microstructure, analogous to $\Omega(\boldsymbol{r})$ ; an assumption not fulfilled in for example brain white matter. This non-zero contribution from magnetic microstructure to $\bar{\Omega}(\mathbf{R})$ engenders a difference to $\Omega(\boldsymbol{r})$, and adds an extra complication to estimating susceptibility in tissue. Using this framework, we derived the mesoscopic Larmor frequency for a population of infinitely long cylinders with orientations drawn from a fiber orientation distribution function (fODF). The mesoscopic contribution $\bar{\Omega}^{\mathrm{Meso}}(\mathbf{R})$ to the average Larmor frequency $\bar{\Omega}(\mathbf{R})$ from this complex magnetic microstructure, was found to depend on the bulk magnetic susceptibility $\bar{\chi}(\mathbf{R})$ and the fODF via its Laplace expansion coefficients $p_{2m}(\mathbf{R})$ given by Eq. (42).

The Lorentz tensor $\mathbf{L}$ for randomly positioned parallel cylinders pointing along $\hat{\boldsymbol{z}}$ is $\mathbf{L}=\zeta\chi\frac{1}{3}\left(\hat{\boldsymbol{z}}\hat{\boldsymbol{z}}^{\mathrm{T}}-\frac{1}{2}\left(\hat{\boldsymbol{y}}\hat{\boldsymbol{y}}^{\mathrm{T}}-\hat{\boldsymbol{x}}\hat{\boldsymbol{x}}^{\mathrm{T}}\right)\right)$. Our result for a parallel cylinder agrees with the GLTA model assuming scalar susceptibility[38] with the only difference being the last term ($-\frac{1}{3}\mathbf{I}$ in Eq. (32)). This is due to the convention of the dipole field $\Upsilon(\boldsymbol{r})$ (cf. Eq. (3)) so that $\mathbf{L}_{\mathrm{GLTA}}=\mathbf{L}+\frac{1}{3}\zeta\chi\,\mathbf{I}$. Our convention ensures $\mathbf{L}=0$ for isotropic media such as uniformly oriented cylinders when $\mathrm{M}(\mathbf{R})$ is chosen to be a sphere. However, we emphasize that the total frequency shift is invariant to whatever convention we choose. If we use the result for a parallel cylinder and assume an overall infinitely long cylindrical sample (with axis parallel to the cylinders), the macroscopic contribution can be shown to be

$\bar{\Omega}^{\text{Macro}} = -\gamma \mathrm{B}_0 \hat{\mathbf{B}}^{\mathrm{T}} \mathbf{L} \hat{\mathbf{B}}$, which cancels the mesoscopic contribution, resulting in a total frequency shift $\bar{\Omega} = 0$, in agreement with previous results[41].

**Simulations**

In simulation (a), the accuracy of incorporating microstructure only in a mesoscopic region surrounding $\mathbf{R}$, when computing $\bar{\Omega}(\mathbf{R})$ (cf. Figure 7) was investigated. Already when the characteristic size of the mesoscopic region was around 10 times the radius of the cylindrical inclusions, the estimated field agreed very well (within 5%) with the ground truth, and is in agreement with previous findings[69]. This was true whether the mesoscopic cavity was spherical, cubical, or cylindrical. This confirms that the sample can be decomposed using any shape for the mesoscopic region, and explicit microstructure need only be accounted for in the vicinity of $\mathbf{R}$ when the microstructure is statistically homogenous. The rapid convergence with the increasing size of the mesoscopic Lorentz cavity suggests a high accuracy when applied to biological tissues such as brain white matter with typical cell sizes of ~10 micrometers relative to the MRI resolution typically exceeding 1 millimeter. A slight underestimation of the frequency was systematically present, which stems from the finite length of cylinders in the simulation, decreasing the magnitude of the eigenvalues[58].

In general, the size of the mesoscopic sphere, necessary to achieve a given accuracy, depends on the structural universality class of the medium and its correlation length, as demonstrated in Figure 8 showing the scaling of the mean squared difference between the model estimate and ground truth. The observed power law decay can be explained from the decay of the structural correlation function $\Gamma^{vv}(\boldsymbol{r}) \sim r^{-(p+d)}$ from inclusions residing outside the mesoscopic sphere (as derived in appendix C). Here $d$ is the microstructural dimension while $p$ denotes the structural exponent[46].

In simulation (b), we computed the MRI Larmor frequency $\bar{\Omega}_{\mathrm{MRI}}$ in a sample with a slowly varying magnetic microstructure. This was done using our full model (Eq. (19)), and compared to a ground truth calculation (Eq. (54)). Figure 9 shows that the standard deviation of the ratio between the two field estimates decreases when the voxel width compared to the size of the cylinders increases and confirms that using the full model (Eq. (19)) on these particular samples can estimate the ground truth MRI Larmor frequency with high accuracy. For example, assuming a radius around 0.5 μm [67], voxels should ideally be larger than around 20 μm, which is well below standard clinical resolution, if standard deviation should be below 5%. However, as shown in simulation (a), the deviation would decrease further for higher volume fractions. In addition, neglecting the mesoscopic contribution and using the elementary dipole field instead of the voxel averaged one when computing the macroscopic contribution resulted in substantial deviations from ground truth. This underscores that we cannot replace the mesoscopic Lorentz sphere with an empty void, as for the molecular Lorentz cavity. By doing so, we neglect an important contribution to the MRI Larmor frequency.

In simulation (c), we tested the accuracy of the mesoscopic dipole tensor $\mathbf{N}/\zeta$ for a population of infinitely long cylinders, Eq. (41), compared to explicitly calculating $\mathbf{N}/\zeta$ using Eq. (16). Our model agreed with our analytical solution, Eq. (41) and previous results for parallel cylinders[41,60]. The degree of orientation dispersion was varied to make the effect of structural anisotropy clear.

**Implications for QSM**

Macroscopic field measurements provide the input data for QSM/STI, through the MR Larmor frequency shift. Currently, these MRI modalities disregard effects associated with microstructure of magnetic inclusions[31] (e.g., myelinated axons), chemical exchange[32] as well as the massive voxel averaging inherent to imaging. It is equivalent to effectively setting the considered effects to zero, which in the case of cylinders with scalar susceptibility is only achieved in the limit of maximal dispersion, the isotropic distribution. Additionally, current cylinder models used to describe microstructural effects of vasculature[43,70] and WM[38,42] assume parallel cylinders, which is hardly the case in most voxels. Assuming an orientation dispersion around 20 degrees (as found in human corpus callosum[71]) decreases $\mathbf{N}$ by approximately 40% compared to fully parallel cylinders. This underscores that the effect of orientation dispersion on phase MRI in cylindrical networks can be substantial and therefore should not be ignored.

**Extending the biophysical model – towards a realistic biophysical model of WM**

The cylinder model presented here incorporates an important microstructural feature into susceptibility estimation, namely the effects of fiber orientation dispersion. This is an important step in modelling complex biological systems and extends state of the art models which assume fully parallel or maximally dispersed cylinders. Equation (41) bridges these two extremes by providing an expression valid for any degree of dispersion. Nevertheless, our model is still a simplified description as we only consider extra-cylindrical spins, while in reality intra-cylindrical compartments typically must be included. For WM for example, intra-cylindrical spins reside in intra-axonal and myelin water bilayers. Furthermore, only a scalar magnetic susceptibility was considered, and while this may be justified for vascular networks, myelin susceptibility anisotropy should be accounted for in a model of WM[30]. Extending the model to include intra-cylindrical compartmentalization, WM susceptibility anisotropy along with additional types of magnetic inclusions, will therefore be investigated in future studies.

*Raison d'etre for modelling magnetic microstructure*

Realistic microstructure such as in WM of course does not consist of long, perfect concentric cylinders, but of fibers with non-circular cross-sections meandering in-between each other[72,73]. One may therefore naturally question if such simple structures can provide a useful model of complex biological environments[54]. However, if we consider Eq. (13), we see that the mesoscopic contribution is given by the integral of the dipole kernel multiplied by the structure-susceptibility correlation function $\Gamma^{\nu\chi}$. As the dipole kernel can be written in terms of the $l = 2$ STF tensors multiplied with $\mathrm{Y}_2^m$ (see Eq. (38)), the mesoscopic contribution is essentially determined by the projection of the correlation function onto $l = 2$ spherical harmonics. This operation clearly erodes the impact of microstructural details on the signal and hence makes identifying the full microstructure ill-posed. Because of this "degeneracy", it makes sense to model complex environments using simple structures, perhaps using spheres and cylinders. Identifying the appropriate model structures is therefore important and developing models should ideally be done in conjunction with simulations on realistic geometries and other types of validation. This is another aim for future studies.

We believe that the framework proposed here, characterizing both mesoscopic and macroscopic field effects on the NMR and MRI Larmor frequency, along with deriving the dependence of orientation dispersion for a population of cylinders will aid in the pursuit of achieving more faithful susceptibility estimations and better biophysical models of nervous tissue.

## Acknowledgments

This study is funded by the Independent Research Fund (grant 8020-00158B). The authors would like to thank PhD Jonas Lynge Olesen for many fruitful discussions.

## Abbreviations

**BOLD:** Blood Oxygenation Level Dependent. **QSM**: Quantitative Susceptibility Mapping. **STI**: Susceptibility Tensor Imaging. **PSF**: Point-Spread-Function. **WM**: White Matter. **dMRI:** Diffusion MRI. **GE**: Gradient-Recalled echo. **fODF:** Fiber Orientation Distribution Function. **STF**: Symmetric Trace Free Tensor. **SO(3)**: Group of all Rotations about the Origin in Three Dimensional Euclidian Space $\mathbb{R}^3$.

## Appendix A

In this appendix we derive how the NMR, $\bar{\Omega}$, and MRI measured Larmor frequency, $\bar{\Omega}_{\text{MRI}}(\mathbf{R})$, can be expressed by the coarse-grained Larmor frequency $\bar{\Omega}(\mathbf{R})$.

### A1) NMR Larmor frequency $\bar{\Omega}$

The common method for probing $\Omega(\boldsymbol{r})$ in biological tissue is the FID signal, $S(t)$, due to its sensitivity to the local magnetic field pertubations. Assuming $S(t)$ is measured in either the static or diffusion narrowing regime, only the first signal cumulant depends linearly on $t$:

$$S(t) = \exp\left(-it\frac{1}{(1-\zeta)\mathrm{V}}\int d\boldsymbol{r}\left(1-v(\boldsymbol{r})\right)\Omega(\boldsymbol{r}) + O\left(t^2\right)\right). \tag{46}$$

Here we omit the irreversible molecular relaxation and external field inhomogeneities for notational simplicity. The factor $(1-\zeta)$ is the volume fraction of the reporting NMR fluid in the whole sample, while $\zeta$ is the volume fraction of inclusions. The NMR Larmor frequency $\bar{\Omega}$ is then defined by the linear contribution to a polynomial fit of the signal phase

$$\bar{\Omega} = \frac{1}{(1-\zeta)\mathrm{V}} \int d\boldsymbol{r} \left(1 - v(\boldsymbol{r})\right) \Omega(\boldsymbol{r}). \tag{47}$$

Equation (45) describes an integral over the macroscopic sample, while characterized by microscopic quantities $v(\boldsymbol{r})$ and $\Omega(\boldsymbol{r})$. In view of Eq. (6), this separation of scales enables us to coarse-grain the integrand of Eq. (45) on the mesoscale by convolving with the mesoscopic sphere $\mathrm{M}(\mathbf{R})$

$$\begin{aligned} \bar{\Omega} &\approx \frac{1}{(1-\zeta)\mathrm{V}} \int d\mathbf{R} \frac{1}{|\mathrm{M}|} \int_{\mathrm{M}(\mathbf{R})} d\boldsymbol{r} \left(1 - v(\boldsymbol{r})\right) \Omega(\boldsymbol{r}) \\ &= \frac{1}{(1-\zeta)\mathrm{V}} \int d\mathbf{R} \left(1 - \bar{\zeta}(\mathbf{R})\right) \bar{\Omega}(\mathbf{R}). \end{aligned} \tag{48}$$

Using Eqs. (6) and (7) for $\bar{\Omega}(\mathbf{R})$ and $\bar{\zeta}(\mathbf{R})$, and Eqs. (10) and (12) for $\bar{\Omega}^{\mathrm{Macro}}(\mathbf{R})$ and $\bar{\Omega}^{\mathrm{Meso}}(\mathbf{R})$, respectively, we get for the NMR Larmor frequency $\bar{\Omega}$, Eq. (46),

$$\bar{\Omega} = \bar{\Omega}^{\mathrm{Meso}} + \bar{\Omega}^{\mathrm{Macro}} + \bar{\Omega}^{\mathrm{W}}, \tag{49}$$

where

$$\bar{\Omega}^{\mathrm{Meso}} = \gamma \mathrm{B}_0 \hat{\mathbf{B}}^{\mathrm{T}} \frac{1}{(1-\zeta)\mathrm{V}} \int d\mathbf{R} \left(1 - \bar{\zeta}(\mathbf{R})\right) \mathbf{L}(\mathbf{R}) \hat{\mathbf{B}}, \tag{50}$$

$$\bar{\Omega}^{\mathrm{Macro}} = \gamma \mathrm{B}_0 \hat{\mathbf{B}}^{\mathrm{T}} \frac{1}{(1-\zeta)\mathrm{V}} \int d\mathbf{R} \left(1 - \bar{\zeta}(\mathbf{R})\right) \int d\mathbf{R}' \, \Upsilon(\mathbf{R}-\mathbf{R}') \bar{\chi}(\mathbf{R}') \hat{\mathbf{B}}, \tag{51}$$

$$\bar{\Omega}^{\mathrm{W}} = \gamma \mathrm{B}_0 \chi^{\mathrm{W}} \hat{\mathbf{B}}^{\mathrm{T}} \mathbf{N}^{\mathrm{Macro}} \hat{\mathbf{B}}. \tag{52}$$

where the $\mathbf{N}^{\mathrm{Macro}}$ is the sample-specific Lorentz-corrected magnetometric demagnetization tensor[51,52]

$$\mathbf{N}^{\mathrm{Macro}} = \frac{1}{\mathrm{V}} \int d\mathbf{R} \int d\mathbf{R}' \, \Upsilon(\mathbf{R}-\mathbf{R}'), \tag{53}$$

and $(1-\zeta)\mathrm{V}$ acts as a normalization of the sample fluid content. In the case of a statistically homogenous magnetic microstructure in the whole sample, i.e. $\bar{\zeta}(\mathbf{R})=\zeta$ and $\bar{\chi}(\mathbf{R}')=\bar{\chi}$, Eq. (47) reduces to

$$\bar{\Omega}=\gamma \mathrm{B}_0 \hat{\mathbf{B}}^{\mathrm{T}}\left(\mathbf{L}+\mathbf{N}^{\mathrm{Macro}}\left(\bar{\chi}+\chi^{\mathrm{W}}\right)\right)\hat{\mathbf{B}}, \tag{54}$$

Figure 11 visualizes the decomposition of the NMR Larmor frequency $\bar{\Omega}$ into mesoscopic, $\bar{\Omega}^{\mathrm{Meso}}$, and macroscopic contributions, $\bar{\Omega}^{\mathrm{Macro}}$, when the whole sample contains a statistically homogenous magnetic microstructure.

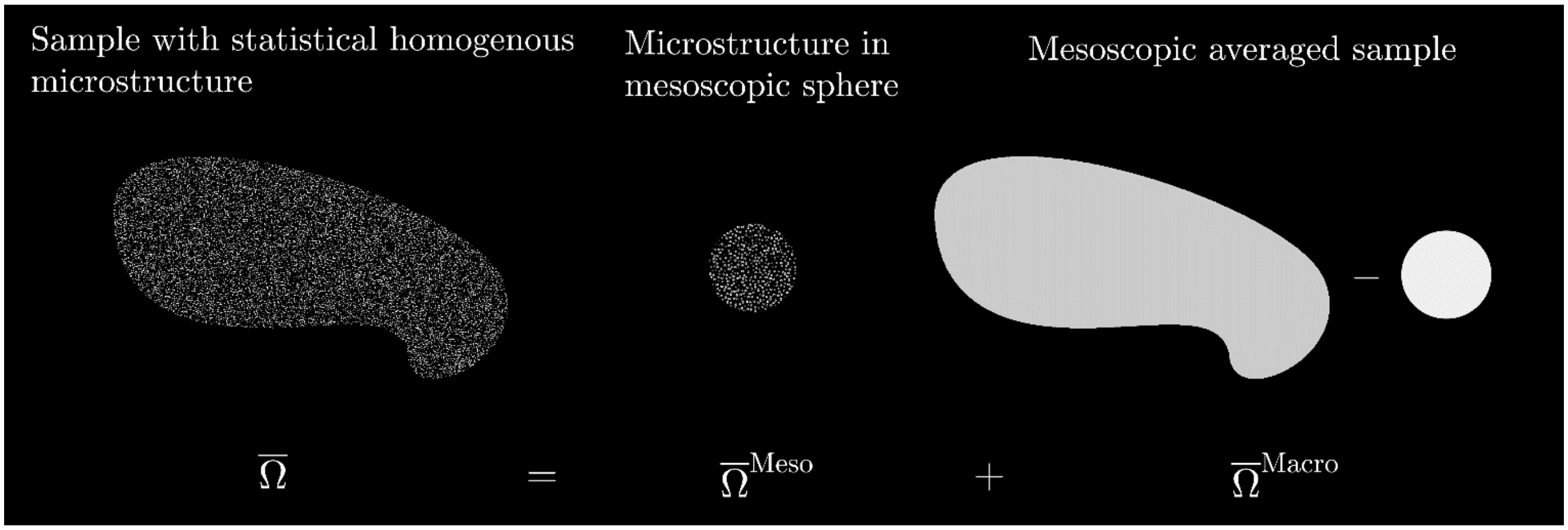

*Figure 11 - Conceptualization of a sample with a given statistically homogenous microstructure, much smaller than the sample size. The measured NMR Larmor frequency $\bar{\Omega}$ (Eq. (45)) of the sample is described by two contributions: $\bar{\Omega}^{\mathrm{Meso}}$ accounts for the Larmor frequency within each mesoscopic sphere $\mathbf{M}$ with explicit microstructure. $\bar{\Omega}^{\mathrm{Macro}}$ describes the Larmor frequency at distances longer than the size of the mesoscopic sphere. We have depicted the deduction of a mesoscopic sphere in $\bar{\Omega}^{\mathrm{Macro}}$, but due the form convention of dipole field (cf. Eq. (3)) and by using a mesoscopic sphere, this contribution is zero. $\bar{\Omega}^{\mathrm{Macro}}$ depends on the coarse-grained magnetic susceptibility $\bar{\chi}$ and the dipole field averaged over the sample $\mathbf{N}^{\mathrm{Macro}}$.*

## A2) MRI Larmor frequency $\bar{\Omega}_{\mathrm{MRI}}(\mathbf{R})$

We now consider the MRI measured Larmor frequency which we denote $\bar{\Omega}_{\mathrm{MRI}}(\mathbf{R})$ to distinguish it from NMR and underscore the effect of imaging. We consider the GE signal, $S(\boldsymbol{k};\mathrm{T_E})$, as this basically samples the FID signal under the influence of image gradients. We consider an ideal MRI acquisition with finite and discrete k-space sampling, and assume the read-out time is sufficiently short, so distortion and blurring of the signal due to time-dependent local phase shifts and relaxation during read-out is negligible. As with the NMR signal, we omit relaxation for notational simplicity and field inhomogeneities from external sources, assuming they can be removed during pre-processing[74]. Our goal is to find out if the error in the measurable Larmor frequency is higher than the usual one introduced by the macroscopic MRI resolution when estimating, e.g., relaxation or diffusion parameters.

Similar to the FID signal, Eq. (44), we may coarse grain $S(\boldsymbol{k};\mathrm{T_E})$, as $\boldsymbol{k}$ is sampled at a scale whose reciprocal determines the imaging resolution, i.e. the macroscopic scale. The normalized signal $S(\mathbf{R};\mathrm{T_E})$ at discrete sample positions $\mathbf{R}$ in image space, measured in the static or diffusion narrowing regime at an echo time $\mathrm{T_E}$ is similar to the FID signal

$$S(\mathbf{R};\mathrm{T_E}) \approx \exp\left(-i\mathrm{T_E}\frac{1}{\left[\mathrm{P}\otimes(1-\bar{\zeta})\right](\mathbf{R})}\left[\mathrm{P}\otimes\left[(1-\zeta)\bar{\Omega}\right]\right](\mathbf{R})+O\left(\mathrm{T_E^2}\right)\right). \tag{55}$$

Here $\mathrm{P}(\mathbf{R})$ denotes the sampling PSF. Hence, the measured MRI Larmor frequency, $\bar{\Omega}_{\mathrm{MRI}}(\mathbf{R})$, becomes

$$\bar{\Omega}_{\mathrm{MRI}}(\mathbf{R}) = \frac{1}{\left[\mathrm{P}\otimes(1-\bar{\zeta})\right](\mathbf{R})}\left[\mathrm{P}\otimes\left[(1-\zeta)\bar{\Omega}\right]\right](\mathbf{R}). \tag{56}$$

This expression coincides with the naiive expectation of obtaining $\bar{\Omega}_{\mathrm{MRI}} = \mathrm{P}\otimes\bar{\Omega}$ only in the case of a constant $\bar{\zeta}(\mathbf{R})$. For a slowly varying $\bar{\zeta}(\mathbf{R})$, it becomes a reasonable approximation, less accurate the more rapidly the inclusion density changes. This is an additional bias in $\bar{\Omega}_{\mathrm{MRI}}(\mathbf{R})$ on top of the common error due to the limited MRI resolution. In mathematical terms, Eq. (53) contains an unavoidable convolution whether we operate in image or k-space.

Staying in image space, we have to integrate the quantity of interest with the PSF according to Eq. (53). An obvious aproximation to this integral is taking the sum over the "voxelized" values of $[(1-\bar{\zeta})\bar{\Omega}](\mathbf{R})$. This approximates the macroscopic contribution, Eq. (10), as

$$\bar{\Omega}^{\text{Macro}}(\mathbf{R}) \approx \sum_{\mathbf{R}'} \bar{\Upsilon}(\mathbf{R}-\mathbf{R}')\,\bar{\chi}(\mathbf{R}'), \tag{57}$$

where the new averaged dipole field is

$$\bar{\Upsilon}(\mathbf{R}-\mathbf{R}') = \int_{C(\mathbf{R}')} d\mathbf{R}''\,\Upsilon(\mathbf{R}-\mathbf{R}''). \tag{58}$$

$C(\mathbf{R}')$ is the voxel centered at $\mathbf{R}'$. This ensures that even though we only sample at discrete position $\mathbf{R}'$, we fill the entire integral of Eq. (10).

Collecting the mesoscopic and macroscopic contributions, including the contribution purely from the NMR fluid, Eq. (4), the MRI measured Larmor frequency becomes

$$\bar{\Omega}_{\text{MRI}}(\mathbf{R}) \approx \gamma B_0 \hat{\mathbf{B}}^{\mathrm{T}} \left( \mathbf{L}(\mathbf{R}) + \sum_{\mathbf{R}'} \bar{\Upsilon}(\mathbf{R}-\mathbf{R}')\,\bar{\chi}(\mathbf{R}') + \chi^{W} \mathbf{N}^{W}(\mathbf{R}) \right) \hat{\mathbf{B}}, \text{ (slowly varying microstructure)} \tag{59}$$

Figure 12 conceptualizes computing the MRI Larmor frequency $\bar{\Omega}_{\text{MRI}}(\mathbf{R})$, when the magnetic microstructure varies slowly on the macroscopic scale.

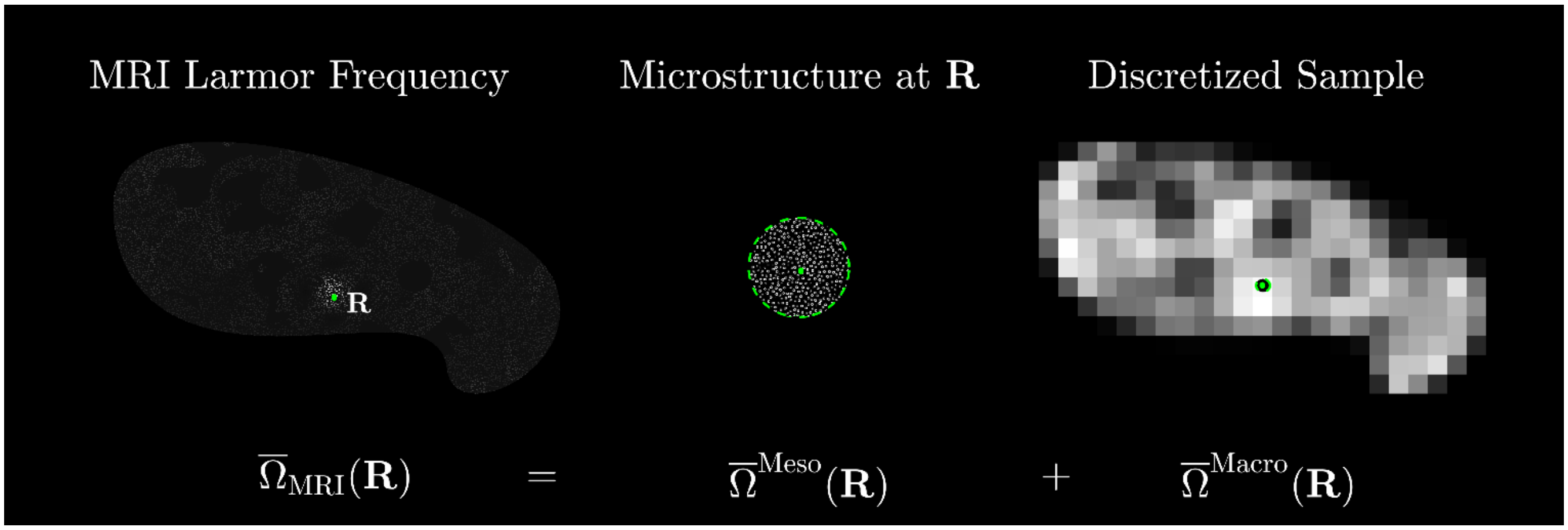

*Figure 12 - Conceptualization of a sample with a varying microstructure, much smaller than the sample size. The measured MRI Larmor frequency $\bar{\Omega}_{\text{MRI}}(\mathbf{R})$ (Eq. (19)) of the sample is described by two contributions: $\bar{\Omega}^{\text{Meso}}(\mathbf{R})$ (Eq. (12)) accounts for the average Larmor frequency from sources within a mesoscopic sphere $M(\mathbf{R})$ with explicit microstructure. $\bar{\Omega}^{\text{Macro}}(\mathbf{R})$ (Eq. (54)) describes the Larmor frequency from sources at distances larger than the size of the mesoscopic sphere. It depends on partitioning the magnetic microstructure into voxels with an associated coarse-grained magnetic susceptibility $\bar{\chi}(\mathbf{R}')$ and the averaged dipole field $\bar{\Upsilon}(\mathbf{R}-\mathbf{R}')$ (Eq. (55)), where $\mathbf{R}'$ corresponds to the sampled points (voxels). $\bar{\Omega}^{\text{Macro}}(0) = 0$ is still zero within this distretization approximation for isotropic voxels and spherical mesoscopic cavity.*

# Appendix B

A full analytical derivation of mesoscopic contribution to the Larmor frequency, Eq. (18), for a population of infinite cylinders with arbitrary orientation dispersion is presented in the following appendices. Integrals are evaluated using the tables in Gradshteyn and Ryzhik[75] and validated numerically and reproduced in supplementary material. References to equations are given as GR(X), where X corresponds to the number of the identity in the original tables.

### B1) Indicator function of a single cylinder

Consider a finite solid cylinder of length $2\mathrm{L}$ and radius $\rho$. Positions $\boldsymbol{r}$ within a cylinder displaced $\boldsymbol{u}$ from the origin can be parametrized by

$$\boldsymbol{r} = \left(u + r\cos\left(\phi\right)\right)\hat{\boldsymbol{u}} + r\sin\left(\phi\right)\hat{\boldsymbol{v}} + s\hat{\boldsymbol{n}}, \tag{60}$$

where $\hat{\boldsymbol{n}}$ is a unit vector along the cylinder axis and $\hat{\boldsymbol{v}}$, $\hat{\boldsymbol{n}}$ and $\hat{\boldsymbol{u}}$ is mutually perpendicular. Hence, ( $r,\phi,s$ ) become local cylinder coordinates. Figure 13 depicts the local coordinates for the cylinder.

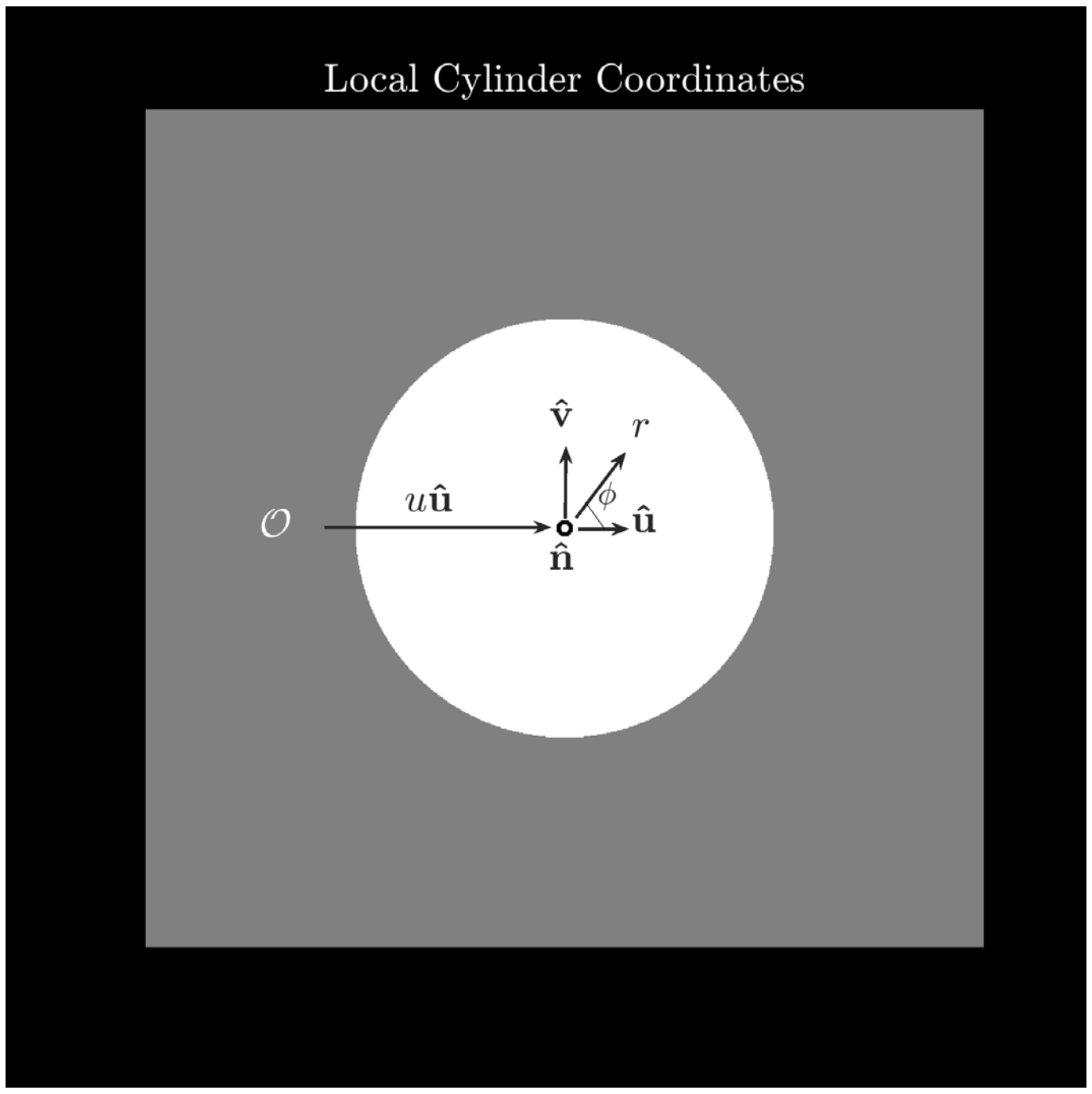

*Figure 13: Local cylinder coordinates. The shortest distance from a given reference point for a fiber is denoted* $u\hat{\boldsymbol{u}}$ *with the axis of the cylinder pointing along* $\hat{\boldsymbol{n}}$ *(perpendicular to the page in the figure).*

The indicator function for a single solid cylinder becomes

$$v(\boldsymbol{r}) = \int_{-L}^{L} ds \int_{0}^{\rho} drr \int_{0}^{2\pi} d\phi\, \delta\left(\boldsymbol{r} - \left(u + r\cos(\phi)\right)\hat{\boldsymbol{u}} - r\sin(\phi)\hat{\boldsymbol{v}} - s\hat{\boldsymbol{n}}\right), \tag{61}$$

and in Fourier-space

$$v(\boldsymbol{k}) = e^{iu\boldsymbol{k}\cdot\hat{\boldsymbol{u}}} \int_{-L}^{L} ds\, e^{isk\hat{\boldsymbol{k}}\cdot\hat{\boldsymbol{n}}} \int_{0}^{\rho} drr \int_{0}^{2\pi} d\phi\, e^{irk\left(\cos(\phi)\,\hat{\boldsymbol{k}}\cdot\hat{\boldsymbol{u}} + \sin(\phi)\hat{\boldsymbol{k}}\cdot\hat{\boldsymbol{v}}\right)}. \tag{62}$$

Performing the radial and angular integration in Eq. (59), using the identities GR (3.338.4)-(6.521.1), yields

$$\int_{0}^{2\pi} d\phi\, e^{irk\left(\cos(\phi)\hat{\boldsymbol{k}}\cdot\hat{\mathbf{u}} + sin(\phi)\hat{\boldsymbol{k}}\cdot\hat{\mathbf{v}}\right)} = 2\pi J_0\left(rk\sqrt{\left(\hat{\boldsymbol{k}}\cdot\hat{\mathbf{u}}\right)^2 + \left(\hat{\boldsymbol{k}}\cdot\hat{\mathbf{v}}\right)^2}\right) \tag{63}$$

and

$$2\pi\int_0^{\rho} dr r J_0\left( rk\sqrt{1-\left(\hat{\boldsymbol{k}}\cdot\hat{\boldsymbol{n}}\right)^2}\right) = \frac{2\pi\rho}{k\sqrt{1-\left(\hat{\boldsymbol{k}}\cdot\hat{\boldsymbol{n}}\right)^2}} J_1\left( \rho k\sqrt{1-\left(\hat{\boldsymbol{k}}\cdot\hat{\boldsymbol{n}}\right)^2}\right), \tag{64}$$

where the identity $\left(\hat{\boldsymbol{k}}\cdot\hat{\boldsymbol{u}}\right)^2+\left(\hat{\boldsymbol{k}}\cdot\hat{\boldsymbol{v}}\right)^2+\left(\hat{\boldsymbol{k}}\cdot\hat{\boldsymbol{n}}\right)^2=1$ was used.

The axial integral in Eq. (59), and its asymptotic limit where $\mathrm{L}\to\infty$, becomes

$$\int_{-\mathrm{L}}^{\mathrm{L}} ds\, e^{isk\boldsymbol{k}\cdot\hat{\boldsymbol{n}}} = \frac{2\mathrm{L}\sin\left(\mathrm{L}k\boldsymbol{k}\cdot\hat{\boldsymbol{n}}\right)}{\mathrm{L}k\boldsymbol{k}\cdot\hat{\boldsymbol{n}}} \to 2\pi\delta\left(\boldsymbol{k}\cdot\hat{\boldsymbol{n}}\right),\ \mathrm{L}\to\infty. \tag{65}$$

Using Eqs. (60)-(62) the Fourier space indicator function for a cylinder of finite length becomes

$$v\left(\boldsymbol{k}\right) = e^{iu\boldsymbol{k}\cdot\hat{\boldsymbol{u}}}\frac{4\pi\rho}{k^2\left(\hat{\boldsymbol{k}}\cdot\hat{\boldsymbol{n}}\right)\sqrt{1-\left(\hat{\boldsymbol{k}}\cdot\hat{\boldsymbol{n}}\right)^2}} J_1\left( \rho k\sqrt{1-\left(\hat{\boldsymbol{k}}\cdot\hat{\boldsymbol{n}}\right)^2}\right)\sin\left(\mathrm{L}k\hat{\boldsymbol{k}}\cdot\hat{\boldsymbol{n}}\right), \tag{66}$$

while in the asymptotic limit, where the length of the cylinder tends to infinity,

$$\begin{aligned} v\left(\boldsymbol{k}\right) &\to e^{iu\hat{\boldsymbol{k}}\cdot\hat{\boldsymbol{u}}}\frac{4\pi^2\rho}{k}J_1\left(\rho k\right)\delta\left(\boldsymbol{k}\cdot\hat{\boldsymbol{n}}\right) \\ &= e^{iu\hat{\boldsymbol{k}}\cdot\hat{\boldsymbol{u}}}v^{\mathrm{2D}}\left(k\right)\delta\left(\boldsymbol{k}\cdot\hat{\boldsymbol{n}}\right),\ \mathrm{L}\to\infty. \quad \left(\text{infinite solid cylinder}\right) \end{aligned} \tag{67}$$

Here $v^{\mathrm{2D}}\left(k\right)$ defines the indicator function in the 2D plane transverse to the orientation $\hat{\boldsymbol{n}}$. Plugging $v\left(\boldsymbol{k}\right)$ into Eq. (17), the total Fourier space correlation function for the infinite solid cylinder becomes

$$\begin{aligned} \Gamma^{vv}\left(\boldsymbol{k}\right) &= \frac{8\pi^2\zeta_c}{k^2}J_1\left(\rho k\right)^2\delta\left(\boldsymbol{k}\cdot\hat{\boldsymbol{n}}\right) - 4\pi^2\frac{\zeta_c^2\delta\left(k\right)}{k}\delta\left(\boldsymbol{k}\cdot\hat{\boldsymbol{n}}\right). \\ &= 2\pi\,\Gamma^{\mathrm{2D}}\left(k\right)\delta\left(\boldsymbol{k}\cdot\hat{\boldsymbol{n}}\right), \quad \left(\text{solid cylinder}\right) \end{aligned} \tag{68}$$

Where $\zeta_c$ is the cylinder volume fraction, and the delta function in the second term of Eq. (65) was interpreted with respect to the orientation of the cylinder axis

$$\delta\left(\boldsymbol{k}\right) = \delta\left(\boldsymbol{k}\cdot\hat{\boldsymbol{u}}\right)\delta\left(\boldsymbol{k}\cdot\hat{\boldsymbol{v}}\right)\delta\left(\boldsymbol{k}\cdot\hat{\boldsymbol{n}}\right) = \frac{\delta\left(k\right)\delta\left(\boldsymbol{k}\cdot\hat{\boldsymbol{n}}\right)}{2\pi k}. \tag{69}$$

Again, $\Gamma^{\mathrm{2D}}\left(k\right)$ defines the autocorrelation perpendicular to the orientation $\hat{\boldsymbol{n}}$. It is axially symmetric and depends only on the radial distance.

## B2) Autocorrelation contribution from single cylinder

Using the explicit result for the autocorrelation, Eq. (65), the contribution from a single cylinder $\mathbf{N}_m$, Eq. (24), to the mesoscopic demagnetization tensor $\mathbf{N}$, Eq. (16), becomes

$$\mathbf{N}_m = \frac{\zeta_m}{(1-\zeta)}\int \frac{dkd\hat{\boldsymbol{k}}\,k}{(2\pi)^3}\boldsymbol{\Upsilon}(\boldsymbol{k})\left(\frac{8\pi^2}{k^2}J_1(\rho_m k)^2 - 4\pi^2\zeta\frac{\delta(k)}{k}\right)\delta(\boldsymbol{k}\cdot\hat{\boldsymbol{n}}_m). \tag{70}$$

The delta function $\delta(\boldsymbol{k}\cdot\hat{\boldsymbol{n}}_m)$ defines a 2D polar integration perpendicular to the orientation $\hat{\boldsymbol{n}}(m)$ of the m'th cylinder. Using GR(6.574.2), the radial integration of Eq. (67) yields

$$\frac{\zeta_m}{(1-\zeta)}\int \frac{dk}{(2\pi)^2}\left(\frac{8\pi^2}{k}J_1(\rho_m k)^2 - 4\pi^2\zeta\delta(k)\right) = \zeta_m. \tag{71}$$

The angular integral of the dipole kernel in Eq. (67), in the plane perpendicular to the cylinder axis $\hat{\boldsymbol{n}}$, can be computed in the laboratory frame, where the cylinder axis is along $\hat{\boldsymbol{z}}$, and then rotated back to $\hat{\boldsymbol{n}}$ specified by the rotation matrix $\mathbf{Q}(\hat{\boldsymbol{n}})$

$$\int \frac{d\hat{\boldsymbol{k}}}{2\pi}\boldsymbol{\Upsilon}(\hat{\boldsymbol{k}})\delta(\hat{\boldsymbol{k}}\cdot\hat{\boldsymbol{n}}) = \mathbf{Q}(\hat{\boldsymbol{n}})\int \frac{d\hat{\boldsymbol{k}}'}{2\pi}\boldsymbol{\Upsilon}(\hat{\boldsymbol{k}}')\delta(\hat{\boldsymbol{k}}'\cdot\hat{\boldsymbol{z}})\mathbf{Q}^{\mathrm{T}}(\hat{\boldsymbol{n}}). \tag{72}$$

Computing the angular integral in Eq. (69) yields

$$\begin{aligned}\int \frac{d\hat{\boldsymbol{k}}}{2\pi}\boldsymbol{\Upsilon}(\hat{\boldsymbol{k}})\delta(\hat{k}_z) &= \frac{1}{2\pi}\int_0^{2\pi} d\varphi\left(\frac{1}{3}\mathbf{I} - \begin{pmatrix}\cos(\varphi)^2 & \sin(\varphi)\cos(\varphi) & 0\\ \sin(\varphi)\cos(\varphi) & \sin(\varphi)^2 & 0\\ 0 & 0 & 0\end{pmatrix}\right)\\ &= \left(\frac{1}{3}\mathbf{I} - \frac{1}{2}\left(\mathbf{I} - \hat{\boldsymbol{z}}^{\mathrm{T}}\hat{\boldsymbol{z}}\right)\right).\end{aligned} \tag{73}$$

Plugging Eq. (70) into Eq. (69) then yields

$$\mathbf{Q}(\hat{\boldsymbol{n}})\left(\frac{1}{3}\mathbf{I} - \frac{1}{2}\left(\mathbf{I} - \hat{\boldsymbol{z}}^{\mathrm{T}}\hat{\boldsymbol{z}}\right)\right)\mathbf{Q}^{\mathrm{T}}(\hat{\boldsymbol{n}}) = \left(\frac{1}{3}\mathbf{I} - \frac{1}{2}\left(\mathbf{I} - \hat{\boldsymbol{n}}^{\mathrm{T}}\hat{\boldsymbol{n}}\right)\right). \tag{74}$$

Combining Eqs. (68) and (71), the contribution from a single cylinder to the autocorrelation becomes

$$\mathbf{N}_m = \zeta_m \left( \frac{1}{3}\mathbf{I} - \frac{1}{2}\left(\mathbf{I} - \hat{\boldsymbol{n}}_m \hat{\boldsymbol{n}}_m^{\mathrm{T}}\right)\right). \tag{75}$$

## B3) Cross correlation of non-parallel cylinders

The total contribution $\Gamma_{wm}(\boldsymbol{k}) + \Gamma_{mw}(\boldsymbol{k})$ from cross-correlations between two non-parallel cylinders, Eq. (25) , can be found using Eq. (64)

$$\begin{aligned} &\Gamma_{wm}(\boldsymbol{k}) + \Gamma_{mw}(\boldsymbol{k}) = \\ &\quad 2\cos(\boldsymbol{k}\cdot\Delta\boldsymbol{u}_{mw}) \frac{16\pi^4}{k^2} \rho_m \rho_w J_1(\rho_m k) J_1(\rho_w k)\, \delta(\boldsymbol{k}\cdot\hat{\boldsymbol{n}}_m)\, \delta(\boldsymbol{k}\cdot\hat{\boldsymbol{n}}_w), \end{aligned} \tag{76}$$

where $\Delta\boldsymbol{u}_{mw} = \boldsymbol{u}_m - \boldsymbol{u}_w$. Due to $\delta(\boldsymbol{k}\cdot\hat{\boldsymbol{n}}_m)\delta(\boldsymbol{k}\cdot\hat{\boldsymbol{n}}_w)$, the integration of Eq. (73) in Eq. (25) collapses to a line integral along $\hat{\boldsymbol{k}} \propto \pm\hat{\boldsymbol{n}}_m \times \hat{\boldsymbol{n}}_w$. Secondly, as $\Delta\boldsymbol{u}_{mw}$ can be decomposed along $\hat{\boldsymbol{n}}_w$, $\hat{\boldsymbol{n}}_m$ and $\hat{\boldsymbol{n}}_m \times \hat{\boldsymbol{n}}_w$, the delta functions pick out the latter component. This simplifies the cosine in Eq. (73)

$$\cos(\boldsymbol{k}\cdot\Delta\boldsymbol{u}_{mw})\,\delta(\boldsymbol{k}\cdot\hat{\boldsymbol{n}}_m)\,\delta(\boldsymbol{k}\cdot\hat{\boldsymbol{n}}_w) = \cos(\pm k u_{mw})\,\delta(\boldsymbol{k}\cdot\hat{\boldsymbol{n}}_m)\,\delta(\boldsymbol{k}\cdot\hat{\boldsymbol{n}}_w). \tag{77}$$

$u_{mw}$ denotes the shortest distance between the two cylinders. The radial integral in Eq. (25) thus becomes (GR6.573.1)

$$\int dk \cos(\pm k u_{mw}) \frac{1}{k^2} J_1(\rho_m k) J_1(\rho_w k) \propto \int dk \frac{1}{k^{\frac{3}{2}}} J_{-\frac{1}{2}}(k u_{mw}) J_1(\rho_m k) J_1(\rho_w k) = 0, \tag{78}$$

as $u_{mw} > \rho_m + \rho_w$, since cylinders are assumed not to overlap. Therefore, $\mathbf{N}^{\mathrm{Cross}}(\boldsymbol{k}) = 0$.

## B4) Cross correlation of parallel cylinders

For the case of two parallel cylinders $\hat{\boldsymbol{n}}_w = \hat{\boldsymbol{n}}_m \equiv \hat{\boldsymbol{n}}$, the total cross-correlation $\Gamma_{wm}(\boldsymbol{k}) + \Gamma_{mw}(\boldsymbol{k})$ corresponds to the first term of the autocorrelation in Eq. (65) multiplied by the phase factor $2\cos\left(u_{mw} k \hat{\boldsymbol{k}}\cdot\hat{\boldsymbol{r}}_{mw}\right)$

$$\Gamma_{wm}(\boldsymbol{k})+\Gamma_{mw}(\boldsymbol{k})=2\cos\left(u_{mw}\boldsymbol{k}\cdot\hat{\boldsymbol{r}}_{mw}\right)\frac{8\pi^2}{k^2}\zeta_w\frac{\rho_m}{\rho_w}J_1(\rho_m k)J_1(\rho_w k)\delta(\boldsymbol{k}\cdot\hat{\boldsymbol{n}}), \tag{79}$$

where $u_{mw}\hat{\boldsymbol{r}}_{mw}$ denotes the shortest distance between the cylinders. The integral in Eq. (25) is over the 2D plane normal to $\hat{\boldsymbol{n}}$. As a consequence, $\hat{\boldsymbol{r}}_{mw}\cdot\hat{\boldsymbol{k}}$ is not constant. For that reason, it is convenient to initially consider the angular integral in Eq. (25) upon integrating Eq. (76)

$$\int\frac{d\hat{\boldsymbol{k}}}{(2\pi)^3}\Upsilon(\boldsymbol{k})\cos\left(ku_{mw}\hat{\boldsymbol{r}}_{mw}\cdot\hat{\boldsymbol{k}}\right)\delta\left(\hat{\boldsymbol{k}}\cdot\hat{\boldsymbol{n}}\right). \tag{80}$$

To evaluate this, consider two cylinders in the eigenframe, where one is positioned at the origin, while the other is positioned along the $u$-axis a distance $u_{mw}\hat{\boldsymbol{u}}$ from the other. Using GR(3.715.18)-(3.715.21), Eq. (77) becomes

$$\begin{aligned}&\frac{1}{(2\pi)^3}\int d\phi\int d\theta\sin(\theta)\cos\left(\cos(\phi)u_{mw}k\right)\cdot\Upsilon(\boldsymbol{k})\delta\left(\hat{\boldsymbol{k}}\cdot\hat{\boldsymbol{n}}\right)\\&=\frac{1}{(2\pi)^2}\left(\frac{J_1(u_{mw}k)}{u_{mw}k}\left(\hat{\boldsymbol{u}}\hat{\boldsymbol{u}}^T-\hat{\boldsymbol{v}}\hat{\boldsymbol{v}}^T\right)+\frac{1}{3}J_0(u_{mw}k)\left(\hat{\boldsymbol{n}}\hat{\boldsymbol{n}}^T-2\hat{\boldsymbol{u}}\hat{\boldsymbol{u}}^T-\hat{\boldsymbol{v}}\hat{\boldsymbol{v}}^T\right)\right).\end{aligned} \tag{81}$$

Using Eq. (78), the radial integral over the 2D plane in Eq. (25) perpendicular to $\hat{\boldsymbol{n}}$ becomes

$$\frac{1}{(2\pi)^2}\int dk\frac{1}{k}J_1(\rho_m k)J_1(\rho_w k)\left(\frac{J_1(u_{mw}k)}{u_{mw}k}\left(\hat{\boldsymbol{u}}\hat{\boldsymbol{u}}^T-\hat{\boldsymbol{v}}\hat{\boldsymbol{v}}^T\right)+\frac{1}{3}J_0(u_{mw}k)\left(\hat{\boldsymbol{n}}\hat{\boldsymbol{n}}^T-2\hat{\boldsymbol{u}}\hat{\boldsymbol{u}}^T-\hat{\boldsymbol{v}}\hat{\boldsymbol{v}}^T\right)\right). \tag{82}$$

Employing GR(6.573.1)-(6.573.2) with the condition $u_{mw}>\rho_m+\rho_w$, the mutual cross correlation between two parallel cylinders separated a distance $u_{mw}$ along $\hat{\boldsymbol{u}}$, becomes

$$\mathbf{N}_{wm}+\mathbf{N}_{mw}=\frac{\zeta_w}{(1-\zeta)}\frac{\rho_m^2}{u_{mw}^2}\left(\hat{\boldsymbol{u}}\hat{\boldsymbol{u}}^T-\hat{\boldsymbol{v}}\hat{\boldsymbol{v}}^T\right), \tag{83}$$

$\left(\text{Parallel cylinders along } \hat{\boldsymbol{n}} \text{ separated } u_{mw} \text{ along } \hat{\boldsymbol{u}}\right)$.

Rotating Eq. (80) an azimuthal angle $\phi$ along $\hat{\boldsymbol{n}}$, the mesoscopic dipole tensor becomes

$$\mathbf{N}_{wm}+\mathbf{N}_{mw}=\frac{\zeta_w}{(1-\zeta)}\frac{\rho_m^2}{u_{mw}^2}\left(\cos(2\phi)\left(\hat{\boldsymbol{u}}\hat{\boldsymbol{u}}^T-\hat{\boldsymbol{v}}\hat{\boldsymbol{v}}^T\right)+\sin(2\phi)\left(\hat{\boldsymbol{u}}\hat{\boldsymbol{v}}^T+\hat{\boldsymbol{v}}\hat{\boldsymbol{u}}^T\right)\right). \tag{84}$$

(Parallel cylinders along $\hat{\boldsymbol{n}}$)

Considering all the pairwise mutual contributions from cylinders pointing along a given direction $\hat{\boldsymbol{n}}$: assuming the cylinders are randomly positioned, and their size is independent of their orientation, the total cross-correlation vanishes due to the angular dependency ($\phi$) in Eq. (81).

# Appendix C

**Argument for the scaling of the variance of the error for increasing mesoscopic sphere.**

We consider the field at the center of the mesoscopic sphere $\mathrm{M}$ with radius $\mathrm{L_M}$, induced from sources outside the mesoscopic sphere. The total sample $\mathrm{V}$ is characterized by the length $\mathrm{L_V}$. This then neglects the microstructure inside the sphere, which is assumed to not affect the overall scaling.

The total Larmor frequency error $\varepsilon$ at the center ($\boldsymbol{r}=0$) is given as

$$\varepsilon=\int_{\mathrm{L_M}}^{\mathrm{L_V}} d\left(\delta\Omega(\boldsymbol{r})\right),\ \overline{\varepsilon}=0, \tag{85}$$

where

$$\begin{aligned} d\left(\delta\Omega(\boldsymbol{r})\right) &\equiv d\Omega\ (\boldsymbol{r})-d\Omega^{\mathrm{uni}}(\boldsymbol{r}) \\ &=\chi\gamma \mathrm{B}_0 \boldsymbol{\Upsilon}^{\hat{\mathbf{B}}}(\boldsymbol{r})\left(v(\boldsymbol{r})-\zeta_1\right)d\mathbf{r} \end{aligned} \tag{86}$$

is the difference between the explicit and coarse grained Larmor frequency contribution from a volume element d$\mathbf{r}$ a distance $\boldsymbol{r}$ from the center. Here $\boldsymbol{\Upsilon}^{\hat{\mathbf{B}}}(\boldsymbol{r})\equiv\hat{\mathbf{B}}^{\mathrm{T}}\boldsymbol{\Upsilon}(\boldsymbol{r})\hat{\mathbf{B}}\sim\frac{1}{\boldsymbol{r}^3}$, while $\zeta_1$ is the volume fraction of a single type of inclusions. The variance becomes

$$\left\langle\varepsilon^2\right\rangle\propto\left\langle\int_{\mathrm{L_M}}^{\mathrm{L_V}} d\boldsymbol{r}\boldsymbol{\Upsilon}^{\hat{\mathbf{B}}}(\boldsymbol{r})\left(v(\boldsymbol{r})-\zeta_1\right)\int_{\mathrm{L_M}}^{\mathrm{L_V}} d\boldsymbol{r}'\boldsymbol{\Upsilon}^{\hat{\mathbf{B}}}(\boldsymbol{r}')\left(v(\boldsymbol{r}')-\zeta_1\right)\right\rangle, \tag{87}$$

As the only stochastic variable here is the indicator function, we obtain

$$\left\langle \varepsilon^2 \right\rangle \propto \int_{L_M}^{L_V} d\boldsymbol{r} \int_{L_M}^{L_V} d\boldsymbol{r}' \Upsilon^{\hat{\mathbf{B}}}(\boldsymbol{r}) \Upsilon^{\hat{\mathbf{B}}}(\boldsymbol{r}') \left\langle \left(v(\boldsymbol{r}) - \zeta_1\right)\left(v(\boldsymbol{r}') - \zeta_1\right) \right\rangle, \tag{88}$$

The mean of the indicator function can be identified as the correlation function $\Gamma^{vv}(\boldsymbol{r}-\boldsymbol{r}') = \left\langle \left(v(\boldsymbol{r}) - \zeta_1\right)\left(v(\boldsymbol{r}') - \zeta_1\right) \right\rangle$, so

$$\left\langle \varepsilon^2 \right\rangle \propto \int_{L_M}^{L_V} d\boldsymbol{r} \int_{L_M}^{L_V} d\boldsymbol{r}' \Upsilon^{\hat{\mathbf{B}}}(\boldsymbol{r}) \Upsilon^{\hat{\mathbf{B}}}(\boldsymbol{r}') \Gamma^{vv}(\boldsymbol{r}-\boldsymbol{r}'). \tag{89}$$

If we assume to be predominantly sensitive to the tail of the correlation function, then[46] $\Gamma^{vv}(\boldsymbol{r}\text{-}\boldsymbol{r}') \sim \left|\frac{\tilde{r}-\tilde{r}'}{l_c}\right|^{-(p+d)}$, where $l_c$ is the correlation length of the microstructure. Introducing a change in variables $L_M \tilde{\boldsymbol{r}} = \boldsymbol{r}$, $L_M \tilde{\boldsymbol{r}}' = \boldsymbol{r}'$ we arrive at the scaling relation

$$\left\langle \varepsilon^2 \right\rangle \propto \left(\frac{L_M}{l_c}\right)^{-(p+d)} \int_1^{\infty} d\tilde{r} \int_1^{\infty} d\tilde{r}' \frac{1}{\tilde{r}} \frac{1}{\tilde{r}'} \left|\tilde{r}\text{-}\tilde{r}'\right|^{-(p+d)} \sim \left(\frac{L_M}{l_c}\right)^{-(p+d)}. \tag{90}$$

Here we used that $L_V \gg L_M$ and replaced the upper limit with infinity, as the integral converges. Hence, we would expect the scaling to depend on dimensionality $d$ of the microstructure and structural exponent $p$.

# References

1. Ogawa S, Lee TM, Kay AR, Tank DW. Brain magnetic resonance imaging with contrast dependent on blood oxygenation. *Proc Natl Acad Sci U S A*. 1990;87(24):9868-9872. doi:10.1073/PNAS.87.24.9868

2. S O, TM L, AS N, P G. Oxygenation-sensitive contrast in magnetic resonance image of rodent brain at high magnetic fields. *Magn Reson Med*. 1990;14(1):68-78. doi:10.1002/MRM.1910140108

3. Ogawa S, Lee T -M. Magnetic resonance imaging of blood vessels at high fields: In vivo and in vitro measurements and image simulation. *Magn Reson Med*. 1990;16(1):9-18.

doi:10.1002/MRM.1910160103

4. Haacke EM, Xu Y, Cheng YCN, Reichenbach JR. Susceptibility weighted imaging (SWI). *Magn Reson Med*. 2004;52(3):612-618. doi:10.1002/mrm.20198

5. Does MD. Inferring brain tissue composition and microstructure via MR relaxometry. *Neuroimage*. 2018;182:136-148. doi:10.1016/j.neuroimage.2017.12.087

6. Lu H, Ge Y. Quantitative evaluation of oxygenation in venous vessels using T2-Relaxation-Under-Spin-Tagging MRI. *Magn Reson Med*. 2008;60(2):357-363. doi:10.1002/MRM.21627

7. Wehrli FW, Fan AP, Rodgers ZB, Englund EK, Langham MC. Susceptibility-based time-resolved whole-organ and regional tissue oximetry. *NMR Biomed*. 2017;30(4). doi:10.1002/nbm.3495

8. Jain V, Langham MC, Wehrli FW. MRI estimation of global brain oxygen consumption rate. *J Cereb Blood Flow Metab*. 2010;30(9):1598-1607. doi:10.1038/JCBFM.2010.49

9. Menon RS, Ogawa S, Kim SG, et al. Functional brain mapping using magnetic resonance imaging. Signal changes accompanying visual stimulation. *Invest Radiol*. 1992;27 Suppl 2(SUPPL. 2):S47-53. doi:10.1097/00004424-199212002-00009

10. Kwong KK, Belliveau JW, Chesler DA, et al. Dynamic magnetic resonance imaging of human brain activity during primary sensory stimulation. *Proc Natl Acad Sci U S A*. 1992;89(12):5675. doi:10.1073/PNAS.89.12.5675

11. Belliveau JW, Kennedy DN, McKinstry RC, et al. Functional Mapping of the Human Visual Cortex by Magnetic Resonance Imaging. *Science (80- )*. 1991;254(5032):716-719. doi:10.1126/SCIENCE.1948051

12. Bandettini PA, Wong EC, Hinks RS, Tikofsky RS, Hyde JS. Time course EPI of human brain function during task activation. *Magn Reson Med*. 1992;25(2):390-397. doi:10.1002/MRM.1910250220

13. Fukunaga M, Li TQ, Van Gelderen P, et al. Layer-specific variation of iron content in cerebral cortex as a source of MRI contrast. *Proc Natl Acad Sci U S A*. 2010;107(8):3834-3839.

doi:10.1073/pnas.0911177107

14. Hammond KE, Lupo JM, Xu D, et al. Development of a robust method for generating 7.0 T multichannel phase images of the brain with application to normal volunteers and patients with neurological diseases. *Neuroimage*. 2008;39(4):1682-1692. doi:10.1016/J.NEUROIMAGE.2007.10.037

15. Duyn JH, Van Gelderen P, Li TQ, De Zwart JA, Koretsky AP, Fukunaga M. High-field MRI of brain cortical substructure based on signal phase. *Proc Natl Acad Sci U S A*. 2007;104(28):11796-11801. doi:10.1073/PNAS.0610821104/ASSET/98E47EFB-3057-4B46-9370-42FDD1888E60/ASSETS/GRAPHIC/ZPQ02807-6637-M08.JPEG

16. Schäfer A, Wharton S, Gowland P, Bowtell R. Using magnetic field simulation to study susceptibility-related phase contrast in gradient echo MRI. *Neuroimage*. 2009;48(1):126-137. doi:10.1016/J.NEUROIMAGE.2009.05.093

17. Salomir R, de Senneville BD, Moonen CT. A fast calculation method for magnetic field inhomogeneity due to an arbitrary distribution of bulk susceptibility. *Concepts Magn Reson*. 2003;19B(1):26-34. doi:10.1002/cmr.b.10083

18. Marques JP, Bowtell R. Application of a Fourier-based method for rapid calculation of field inhomogeneity due to spatial variation of magnetic susceptibility. *Concepts Magn Reson Part B Magn Reson Eng*. 2005;25B(1):65-78. doi:10.1002/cmr.b.20034

19. Jenkinson M, Wilson JL, Jezzard P. Perturbation method for magnetic field calculations of nonconductive objects. *Magn Reson Med*. 2004;52(3):471-477. doi:10.1002/mrm.20194

20. Wharton S, Schäfer A, Bowtell R. Susceptibility mapping in the human brain using threshold-based k-space division. *Magn Reson Med*. 2010;63(5):1292-1304. doi:10.1002/MRM.22334

21. Liu T, Spincemaille P, De Rochefort L, Kressler B, Wang Y. Calculation of susceptibility through multiple orientation sampling (COSMOS): A method for conditioning the inverse problem from measured magnetic field map to susceptibility source image in MRI. *Magn Reson Med*. 2009;61(1):196-204. doi:10.1002/mrm.21828

22. Li L, Leigh JS. Quantifying arbitrary magnetic susceptibility distributions with MR. *Magn Reson Med*. 2004;51(5):1077-1082. doi:10.1002/MRM.20054

23. Deistung A, Schweser F, Reichenbach JR. Overview of quantitative susceptibility mapping. *NMR Biomed*. 2017;30(4):e3569. doi:10.1002/nbm.3569

24. Wang Y, Liu T. Quantitative susceptibility mapping (QSM): Decoding MRI data for a tissue magnetic biomarker. *Magn Reson Med*. 2015;73(1):82-101. https://onlinelibrary.wiley.com/doi/full/10.1002/mrm.25358. Accessed April 1, 2021.

25. Shmueli K, De Zwart JA, Van Gelderen P, Li TQ, Dodd SJ, Duyn JH. Magnetic susceptibility mapping of brain tissue in vivo using MRI phase data. *Magn Reson Med*. 2009;62(6):1510-1522. doi:10.1002/MRM.22135

26. de Rochefort L, Brown R, Prince MR, Wang Y. Quantitative MR susceptibility mapping using piece-wise constant regularized inversion of the magnetic field. *Magn Reson Med*. 2008;60(4):1003-1009. doi:10.1002/mrm.21710

27. Marques JP, Maddage R, Mlynarik V, Gruetter R. On the origin of the MR image phase contrast: An in vivo MR microscopy study of the rat brain at 14.1 T. *Neuroimage*. 2009;46(2):345-352. doi:10.1016/J.NEUROIMAGE.2009.02.023

28. Liu C. Susceptibility tensor imaging. *Magn Reson Med*. 2010;63(6):1471-1477. doi:10.1002/mrm.22482

29. Lee J, Shmueli K, Fukunaga M, et al. Sensitivity of MRI resonance frequency to the orientation of brain tissue microstructure. *Proc Natl Acad Sci U S A*. 2010;107(11):5130-5135. doi:10.1073/pnas.0910222107

30. Van Gelderen P, Mandelkow H, De Zwart JA, Duyn JH. A torque balance measurement of anisotropy of the magnetic susceptibility in white matter. *Magn Reson Med*. 2015;74(5):1388-1396. doi:10.1002/mrm.25524

31. He X, Yablonskiy DA. Biophysical mechanisms of phase contrast in gradient echo MRI. *Proc Natl Acad Sci U S A*. 2009;106(32):13558-13563. doi:10.1073/pnas.0904899106

32. Zhong K, Leupold J, von Elverfeldt D, Speck O. The molecular basis for gray and white matter contrast in phase imaging. *Neuroimage*. 2008;40(4):1561-1566. doi:10.1016/J.NEUROIMAGE.2008.01.061

33. Yablonskiy DA, Luo J, Sukstanskii AL, Iyer A, Cross AH. Biophysical mechanisms of MRI signal frequency contrast in multiple sclerosis. *Proc Natl Acad Sci U S A*. 2012;109(35):14212-14217. doi:10.1073/pnas.1206037109

34. Ye FQ, Allen PS. Relaxation enhancement of the transverse magnetization of water protons in paramagnetic suspensions of red blood cells. *Magn Reson Med*. 1995;34(5):713-720. doi:10.1002/MRM.1910340510

35. Durrant CJ, Hertzberg MP, Kuchel PW. Magnetic susceptibility: Further insights into macroscopic and microscopic fields and the sphere of Lorentz. *Concepts Magn Reson Part A*. 2003;18A(1):72-95. doi:10.1002/CMR.A.10067

36. Ruh A, Scherer H, Kiselev VG. The larmor frequency shift in magnetically heterogeneous media depends on their mesoscopic structure. *Magn Reson Med*. 2018;79(2):1101-1110. doi:10.1002/mrm.26753

37. Wills AP. The Theory of Electrons and Its Applications to the Phenomena of Light and Radiant Heat . By H. A. Lorentz. . *Science (80- )*. 1910;31(789):221-223. doi:10.1126/SCIENCE.31.789.221.B

38. Yablonskiy DA, Sukstanskii AL. Generalized Lorentzian Tensor Approach (GLTA) as a biophysical background for quantitative susceptibility mapping. *Magn Reson Med*. 2015;73(2):757-764. doi:10.1002/mrm.25538

39. Luo J, He X, Yablonskiy DA. Magnetic susceptibility induced white matter MR signal frequency shifts—experimental comparison between Lorentzian sphere and generalized Lorentzian approaches. *Magn Reson Med*. 2014;71(3):1251-1263. doi:10.1002/MRM.24762

40. Wharton S, Bowtell R. Effects of white matter microstructure on phase and susceptibility maps. *Magn Reson Med*. 2015;73(3):1258-1269. doi:10.1002/mrm.25189

41. Yablonskiy DA, He X, Luo J, Sukstanskii AL. Lorentz sphere versus generalized Lorentzian approach: What would lorentz say about it? *Magn Reson Med*. 2014;72(1):4-7. doi:10.1002/mrm.25230

42. Wharton S, Bowtell R. Fiber orientation-dependent white matter contrast in gradient echo MRI. *Proc Natl Acad Sci U S A*. 2012;109(45):18559-18564. doi:10.1073/pnas.1211075109

43. Yablonskiy DA, Haacke EM. Theory of NMR signal behavior in magnetically inhomogeneous tissues: The static dephasing regime. *Magn Reson Med*. 1994;32(6):749-763. doi:10.1002/mrm.1910320610

44. Sati P, van Gelderen P, Silva AC, et al. Micro-compartment specific T2* relaxation in the brain. *Neuroimage*. 2013;77:268-278. doi:10.1016/j.neuroimage.2013.03.005

45. Kiselev VG. Larmor frequency in heterogeneous media. *J Magn Reson*. 2019;299:168-175. doi:10.1016/j.jmr.2018.12.008

46. Novikov DS, Jensen JH, Helpern JA, Fieremans E. Revealing mesoscopic structural universality with diffusion. *Proc Natl Acad Sci U S A*. 2014;111(14):5088-5093. doi:10.1073/pnas.1316944111

47. Kleban E, Tax CMW, Rudrapatna US, Jones DK, Bowtell R. Strong diffusion gradients allow the separation of intra- and extra-axonal gradient-echo signals in the human brain. *Neuroimage*. 2020;217:116793. doi:10.1016/J.NEUROIMAGE.2020.116793

48. Yablonskiy DA, Sukstanskii AL. Lorentzian effects in magnetic susceptibility mapping of anisotropic biological tissues. *J Magn Reson*. 2018;292:129-136. doi:10.1016/J.JMR.2018.04.014

49. Duyn JH. Studying brain microstructure with magnetic susceptibility contrast at high-field. *Neuroimage*. 2018;168:152-161. doi:10.1016/J.NEUROIMAGE.2017.02.046

50. Ruh A, Kiselev VG. Calculation of Larmor precession frequency in magnetically heterogeneous media. *Concepts Magn Reson Part A*. 2018;47A(1):e21472. doi:10.1002/cmr.a.21472

51. Osborn JA. Demagnetizing Factors of the General Ellipsoid. *Phys Rev*. 1945;67(11-12):351.

doi:10.1103/PhysRev.67.351

52. MOSKOWITZ R, TORRE E DELLA. Theoretical Aspects of Demagnetization Tensors. *IEEE Trans Magn*. 1966;MAG-2(4):739-744. doi:10.1109/TMAG.1966.1065973

53. Novikov DS, Fieremans E, Jespersen SN, Kiselev VG. Quantifying brain microstructure with diffusion MRI: Theory and parameter estimation. *NMR Biomed*. 2019;32(4):e3998. doi:10.1002/nbm.3998

54. Novikov DS, Kiselev VG, Jespersen SN. On modeling. *Magn Reson Med*. 2018;79(6):3172-3193. doi:10.1002/MRM.27101

55. Kiselev VG, Novikov DS. Transverse NMR relaxation in biological tissues. *Neuroimage*. 2018;182:149-168. doi:10.1016/j.neuroimage.2018.06.002

56. Yablonskiy DA, Sukstanskii AL. Biophysical mechanisms of myelin-induced water frequency shifts. *Magn Reson Med*. 2014;71(6):1956-1958. doi:10.1002/mrm.25214

57. Lee J, Shmueli K, Fukunaga M, et al. Sensitivity of MRI resonance frequency to the orientation of brain tissue microstructure. *Proc Natl Acad Sci U S A*. 2010;107(11):5130-5135. doi:10.1073/pnas.0910222107

58. Beleggia M, De Graef M. On the computation of the demagnetization tensor field for an arbitrary particle shape using a Fourier space approach. *J Magn Magn Mater*. 2003;263(1-2). doi:10.1016/S0304-8853(03)00238-5

59. Beleggia M. A fourier-space approach for the computation of magnetostatic interactions between arbitrarily shaped particles. *IEEE Trans Magn*. 2004;40(4 II):2149-2151. doi:10.1109/TMAG.2004.830214

60. Ruh A, Kiselev VG. Larmor frequency dependence on structural anisotropy of magnetically heterogeneous media. *J Magn Reson*. 2019;307:106584. doi:10.1016/j.jmr.2019.106584

61. Caciagli A, Baars RJ, Philipse AP, Kuipers BWM. Exact expression for the magnetic field of a finite cylinder with arbitrary uniform magnetization. *J Magn Magn Mater*. 2018;456:423-432. doi:10.1016/J.JMMM.2018.02.003

62. Fisher NI, Lewis T, Embleton BJJ. *Statistical Analysis of Spherical Data*. Cambridge University Press; 1987. doi:10.1017/CBO9780511623059

63. Thorne KS. Multipole expansions of gravitational radiation. *Rev Mod Phys*. 1980;52(2):299-339. doi:10.1103/RevModPhys.52.299

64. Sandgaard A, Kiselev VG, Shemesh N, Jespersen SN. Rotation-Free Mapping of Magnetic Tissue Properties in White Matter. In: *ISMRM*. ; 2021.

65. Novikov DS, Veraart J, Jelescu IO, Fieremans E. Rotationally-invariant mapping of scalar and orientational metrics of neuronal microstructure with diffusion MRI. *Neuroimage*. 2018;174:518-538. doi:10.1016/J.NEUROIMAGE.2018.03.006

66. Jackson 1925-2016 JD. *Classical Electrodynamics*. Third edition. New York : Wiley, [1999] ©1999 https://search.library.wisc.edu/catalog/999849741702121.

67. Aboitiz F, Scheibel AB, Fisher RS, Zaidel E. Fiber composition of the human corpus callosum. *Brain Res*. 1992;598(1-2):143-153. doi:10.1016/0006-8993(92)90178-C

68. Lu X, Ma Y, Chang EY, et al. Simultaneous Quantitative Susceptibility Mapping (QSM) and R2* for High Iron Concentration Quantification with Three-Dimensional Ultrashort Echo Time (UTE) Sequences – an Echo Dependence Study. *Magn Reson Med*. 2018;79(4):2315. doi:10.1002/MRM.27062

69. Aravamudhan S. Magnetized materials: Contributions inside Lorentz ellipsoids. *Indian J Phys*. 2005;79(9).

70. Kiselev VG, Posse S. Analytical Model of Susceptibility-Induced MR Signal Dephasing: Effect of Diffusion in a Microvascular Network. *Magn Reson Med*. 1999;41:499-509. doi:10.1002/(SICI)1522-2594(199903)41:3

71. Ronen I, Budde M, Ercan E, Annese J, Techawiboonwong A, Webb A. Microstructural organization of axons in the human corpus callosum quantified by diffusion-weighted magnetic resonance spectroscopy of N-acetylaspartate and post-mortem histology. *Brain Struct Funct*. 2014;219(5):1773-1785. doi:10.1007/s00429-013-0600-0

72. Lee HH, Yaros K, Veraart J, et al. Along-axon diameter variation and axonal orientation dispersion revealed with 3D electron microscopy: implications for quantifying brain white matter microstructure with histology and diffusion MRI. *Brain Struct Funct*. 2019;224(4):1469-1488. doi:10.1007/s00429-019-01844-6

73. Andersson M, Kjer HM, Rafael-Patino J, et al. Axon morphology is modulated by the local environment and impacts the noninvasive investigation of its structure–function relationship. *Proc Natl Acad Sci U S A*. 2021;117(52):33649-33659. doi:10.1073/PNAS.2012533117

74. Zhou D, Liu T, Spincemaille P, Wang Y. Background field removal by solving the Laplacian boundary value problem. *NMR Biomed*. 2014;27(3):312-319. doi:10.1002/nbm.3064

75. Zwillinger D. *Table of Integrals, Series, and Products: Eighth Edition*. Elsevier Inc.; 2014. doi:10.1016/C2010-0-64839-5